\documentclass[12pt]{article}
\usepackage{amsmath}
\usepackage{amsfonts}
\usepackage{amssymb}
\usepackage{graphicx}
\providecommand{\U}[1]{\protect\rule{.1in}{.1in}}
\providecommand{\U}[1]{\protect\rule{.1in}{.1in}}
\providecommand{\U}[1]{\protect\rule{.1in}{.1in}}
\providecommand{\U}[1]{\protect\rule{.1in}{.1in}}
\providecommand{\U}[1]{\protect\rule{.1in}{.1in}}
\providecommand{\U}[1]{\protect\rule{.1in}{.1in}}
\providecommand{\U}[1]{\protect\rule{.1in}{.1in}}
\providecommand{\U}[1]{\protect\rule{.1in}{.1in}} \textwidth17.1cm \oddsidemargin -0.5cm

\newcommand{\be}{\begin{equation}}
\newcommand{\ee}{\end{equation}}
\newcommand{\ben}{\begin{equation*}}
\newcommand{\een}{\end{equation*}}
\newcommand{\ar}{\begin{array}}
\newcommand{\arn}{\end{array}}

\def\pnot{\mbox{${\not{\hbox{\kern-3.0pt$p$}}}$}}
\def\qnot{\mbox{${\not{\hbox{\kern-2.0pt$q$}}}$}}
\def\enot{\mbox{${\not{\hbox{\kern-2.0pt$e$}}}$}}
\def\knot{\mbox{${\not{\hbox{\kern-2.0pt$k$}}}$}}

\def\fun#1#2{\lower3.6pt\vbox{\baselineskip0pt\lineskip.9pt\ialign
{$\mathsurround=0pt#1\hfil##\hfil$\crcr#2\crcr\sim\crcr}}}

\begin{document}

\begin{titlepage}

\begin{center}
{\bf The dipole form of the gluon part of the BFKL kernel$^{~\ast}$}
\end{center}
\vskip 0.5cm \centerline{V.S.~Fadin$^{a\,\dag}$,
R.~Fiore$^{b\,\ddag}$, A.V.~Grabovsky$^{a\,\dag\dag}$, A.
Papa$^{b\,\ddag\dag}$} \vskip .6cm

\centerline{\sl $^{a}$ Budker Institute of Nuclear Physics, 630090
Novosibirsk, Russia} \centerline{\sl Novosibirsk State University,
630090 Novosibirsk, Russia} \centerline{\sl $^{b}$ Dipartimento di
Fisica, Universit\`a della Calabria,} \centerline{\sl Istituto
Nazionale di Fisica Nucleare, Gruppo collegato di Cosenza,}
\centerline{\sl Arcavacata di Rende, I-87036 Cosenza, Italy}
\vskip 2cm

\begin{abstract}
The dipole form of the gluon part of the colour singlet BFKL
kernel in the next-to-leading order (NLO) is obtained in the
coordinate representation by direct transfer from the momentum
representation, where the kernel was calculated before.  With this
paper the transformation of the NLO BFKL kernel to the dipole
form, started a few months ago with the quark part of the kernel,
is completed.

\end{abstract}
\vfill \hrule \vskip.3cm \noindent $^{\ast}${\it Work supported in
part by the RFBR grant 07-02-00953, in part by the RFBR--MSTI
grant 06-02-72041 and in part by Ministero Italiano
dell'Istruzione, dell'Universit\`a e della Ricerca.} \vfill $
\begin{array}{ll} ^{\dag}\mbox{{\it e-mail address:}} &
\mbox{FADIN@INP.NSK.SU}\\
^{\ddag}\mbox{{\it e-mail address:}} &
\mbox{FIORE@CS.INFN.IT}\\
^{\dag\dag}\mbox{{\it e-mail address:}} &
\mbox{A.V.GRABOVSKY@INP.NSK.SU}\\
^{\ddag\dag}\mbox{{\it e-mail address:}} &
\mbox{PAPA@CS.INFN.IT}\\
\end{array}
$

\end{titlepage}

\vfill \eject

\section{Introduction}

The BFKL equation~\cite{BFKL} allows to find the Green's function for the
scattering of two Reggeized gluons, which determines the high-energy behaviour
of QCD amplitudes. It is an integral equation in the space of momenta
transverse to the momenta of colliding particles. The kernel of the BFKL
equation is known now in the next-to-leading order (NLO) both for the forward
scattering~\cite{FL98}, i.e. for $t=0$ and color singlet (Pomeron) in the
$t$-channel, and for any fixed (not growing with energy) momentum transfer $t$
and any possible color state in the $t$-channel~\cite{FFP99,FG00,FF05}.

The Pomeron channel is the most important for phenomenological
applications. However, from the theoretical point of view the
color octet case seems to be even more important because of the
gluon Reggeization. Indeed, high-energy QCD can be reformulated in
terms of a gauge-invariant effective field theory for Reggeized
gluon interactions~\cite{Lipatov:1995pn}, so that the primary
Reggeon in QCD is not the Pomeron, but the Reggeized gluon. The main
property of the octet BFKL kernel which is required by the
Reggeization is that the gluon trajectory turns out to be its
eigenfunction. This property can be verified in the momentum
representation, i.e. the representation in which the BFKL approach
was originally developed. On the contrary, a remarkable property
of the colour singlet BFKL kernel in the leading
approximation~\cite{Lipatov:1985uk} is entirely concerned with the
coordinate representation. It is the famous conformal invariance,
which is extremely important for finding solutions of the
equation.

In the NLO the coordinate representation of the colour singlet
BFKL kernel is also very interesting. First, it reveals conformal
properties of the kernel. Evidently, the conformal invariance is
violated by renormalization. One may wonder,  however, whether the
renormalization is the only source of the violation. If so, one
can expect the conformal invariance of the NLO BFKL kernel in
supersymmetric extensions of QCD.

Another important reason for transferring the colour singlet NLO BFKL kernel
in the coordinate representation is to make possible the comparison with the
color dipole approach for high-energy processes~\cite{dipole}. This approach
gives a clear physical picture of high-energy processes and can be naturally
extended from the regime of low parton densities to the saturation
regime~\cite{GLR83}, where the evolution equations of parton densities with
energy become nonlinear. In general, there is an infinite hierarchy of coupled
equations~\cite{Balitsky,CGC}. In the simplest case of a large nucleus as a
target, this set of equations is reduced to the BK (Balitsky-Kovchegov)
equation~\cite{Balitsky}.

It is affirmed~\cite{dipole,Balitsky} that in the linear regime the color
dipole framework gives the same results as the BFKL one for the color singlet
channel. Before the advent of the dipole approach, the leading order color
singlet BFKL kernel was investigated in the coordinate representation in
detail in Ref.~\cite{Lipatov:1985uk}. More recently, the relation between BFKL
and color dipole approaches was analyzed in the leading order in
Ref.~\cite{Bartels:2004ef}. The extension of the analysis to the NLO was
started a few months ago. In Ref.~\cite{Kovchegov:2006wf} the quark contribution
to the color dipole approach at large number of colors $N_{c}$ has been
transferred from the coordinate to the momentum representation and it has been
verified that the resulting contribution to the NLO Pomeron intercept agrees
with the well-known result of Ref.~\cite{FL98}. In Refs.~\cite{Fadin:2006ha,
Fadin:2007ee} the \textquotedblleft non-Abelian\textquotedblright\ (leading in
$N_{c}$) and \textquotedblleft Abelian\textquotedblright\ parts of the quark
contribution to the non-forward BFKL kernel in the momentum
representation~\cite{FFP99} have been transformed to the coordinate one and it
has been found that the dipole form of the quark contribution agrees with the
result obtained in Ref.~\cite{Balitsky:2006wa} by the direct
calculation of the quark contribution to the dipole kernel in the coordinate
representation.

Evidently, the main and the most important part of the BFKL kernel is given by
the gluon contribution. The aim of this work is to consider this part of the
kernel in the NLO and to find its dipole form.

The paper is organized as follows: in Section~\ref{sec:notation} we give basic
definitions, fix our notations and recall the main results of
the papers in Refs.~\cite{Fadin:2006ha, Fadin:2007ee}; in Section~\ref{sec:decomposition}
we describe the decomposition of the kernel into \textquotedblleft
planar\textquotedblright\ and \textquotedblleft symmetric\textquotedblright
\ parts; in Section~\ref{sec:trajectory} we present the NLO gluon trajectory
in a convenient form; in Section~\ref{sec:octet} we show a convenient
representation for the real part of the colour octet kernel entering into the
\textquotedblleft planar\textquotedblright\ part; in
Section~\ref{sec:cancellation} we discuss the cancellation of the infrared
singularities and give the resulting form of the colour octet kernel; in
Section~\ref{sec:symmetric part} we give the \textquotedblleft
symmetric\textquotedblright\ part of the kernel in a form suitable for the
subsequent transformation; in Section~\ref{sec:Dipole form} we present the
general structure of the dipole form of the kernel; in
Section~\ref{sec:planar} we describe the procedure to transfer the
\textquotedblleft planar\textquotedblright\ piece of the gluon part of the
kernel from the momentum to the coordinate representation; in
Section~\ref{sec:symmetric} we outline this procedure for the
\textquotedblleft symmetric\textquotedblright\ piece; in Section
\ref{sec:result}\ we present our final result; in Section~\ref{sec:conclusion}
we draw our conclusions. Appendices contain all necessary integrals.

\section{Basic definitions and notation}

\label{sec:notation}

We use the same notation as in Ref.~\cite{Fadin:2006ha}: $\vec{q}
_{i}^{\;\prime}$ and $\vec{q}_{i}$, $i=1,2$, represent the transverse momenta
of Reggeons in initial and final $t$-channel states, while $\vec{r}
_{i}^{\;\prime}$ and $\vec{r}_{i}$ are the corresponding conjugate
coordinates. The state normalization is
\begin{equation}
\langle\vec{q}|\vec{q}^{\;\prime}\rangle=\delta(\vec{q}-\vec{q}^{\;\prime
})\;,\;\;\;\;\;\langle\vec{r}|\vec{r}^{\;\prime}\rangle=\delta(\vec{r}-\vec
{r}^{\;\prime})\;, \label{normalization}
\end{equation}
so that
\begin{equation}
\langle\vec{r}|\vec{q}\rangle=\frac{e^{i\vec{q}\,\vec{r}}}{(2\pi)^{1+\epsilon
}}\;,
\end{equation}
where $\epsilon=(D-4)/2$; $D-2$ is the dimension of the transverse space and
it is taken different from $2$ for the regularization of divergences. We will
also use the notation $\vec{q}=\vec{q}_{1}+\vec{q}_{2},\;\;\vec{q}^{\;\prime
}=\vec{q}_{1}^{\;\prime}+\vec{q}_{2}^{\;\prime};\;\;\vec{k}=\vec{q}_{1}
-\vec{q}_{1}^{\;\prime}=\vec{q}_{2}^{\;\prime}-\vec{q}_{2}$ and for brevity we
will usually write $\vec{p}_{ij^{\prime}}=\vec{p}_{i}-\vec{p}_{j}^{\;\prime}$.

The BFKL kernel in the operator form is written as
\begin{equation}
\hat{\mathcal{K}}=\hat{\omega}_{1}+\hat{\omega}_{2}+\hat{\mathcal{K}}_{r}\;,
\label{operator of the BFKL kernel}
\end{equation}
where
\begin{equation}
\langle\vec{q}_{i}|\hat{\omega}_{i}|\vec{q}_{i}^{\;\prime}\rangle=\delta
(\vec{q}_{ii^{\prime}})\omega(-\vec{q}_{i}^{\;2})\;, \label{trajectory ff}
\end{equation}
with $\omega(t)$ the gluon Regge trajectory, and $\hat{\mathcal{K}}_{r}$
represents real particle production in Reggeon collisions. The $s$-channel
discontinuities of scattering amplitudes for the processes $A+B\rightarrow
A^{\prime}+B^{\prime}$ have the form
\begin{equation}
-4i(2\pi)^{D-2}\delta(\vec{q}_{A}-\vec{q}_{B})\mbox{disc}_{s}\mathcal{A}
_{AB}^{A^{\prime}B^{\prime}}=\langle A^{\prime}\bar{A}|e^{Y\hat{\mathcal{K}}
}\frac{1}{\hat{\vec{q}}_{1}^{\;2}\hat{\vec{q}}_{2}^{\;2}}|\bar{B}^{\prime
}B\rangle\;. \label{discontinuity representation}
\end{equation}
In this equation $Y=\ln(s/s_{0})$, $s_{0}$ is an appropriate energy scale,
$\;\;q_{A}=p_{A^{\prime}}-p_{A},\;\;q_{B}=p_{B}-p_{B^{\prime}}$, and
\begin{equation}
\langle\vec{q}_{1},\vec{q}_{2}|\hat{\mathcal{K}}|\vec{q}_{1}^{\;\prime}
,\vec{q}_{2}^{\;\prime}\rangle=\delta(\vec{q}-\vec{q}^{\;\prime})\frac{1}
{\vec{q}_{1}^{\;2}\vec{q}_{2}^{\;2}}\mathcal{K}(\vec{q}_{1},\vec{q}
_{1}^{\;\prime};\vec{q})\;, \label{kernel ff}
\end{equation}
\begin{equation}
\langle\vec{q}_{1},\vec{q}_{2}|\bar{B}^{\prime}B\rangle=4p_{B}^{-}\delta
(\vec{q}_{B}-\vec{q}_{1}-\vec{q}_{2}){\Phi}_{B^{\prime}B}(\vec{q}_{1},\vec
{q}_{2})\;, \label{impact BB}
\end{equation}
\begin{equation}
\langle A^{\prime}\bar{A}|\vec{q}_{1},\vec{q}_{2}\rangle=4p_{A}^{+}\delta
(\vec{q}_{A}-\vec{q}_{1}-\vec{q}_{2}){\Phi}_{A^{\prime}A}(\vec{q}_{1},\vec
{q}_{2})\;, \label{impact AA}
\end{equation}
where $p^{\pm}=(p_{0}\pm p_{z})/\sqrt{2}$; the kernel $\mathcal{K}(\vec{q}
_{1},\vec{q}_{1}^{\;\prime};\vec{q})$ and the impact factors $\Phi$ are
expressed through the Reggeon vertices according to Ref.~\cite{FF98}. Note
that the appearance of the factors $(\hat{\vec{q}}_{1}^{\;2}\hat{\vec{q}}
_{2}^{\;2})^{-1}$ in (\ref{discontinuity representation}) and $(\vec{q}
_{1}^{\;2}\vec{q}_{2}^{\;2})^{-1}$ in (\ref{kernel ff}) cannot be explained by
a change of the normalization~(\ref{normalization}). We have used a freedom in
the definition of the kernel. Indeed, one can change the form of the kernel
(in any representation) performing the operator transformation
\begin{equation}
\hat{\mathcal{K}}\rightarrow\hat{\mathcal{O}}^{-1}\hat{\mathcal{K}}
\hat{\mathcal{O}}~,\;\;\langle A^{\prime}\bar{A}|\rightarrow\langle A^{\prime
}\bar{A}|\hat{\mathcal{O}}~,\;\;\frac{1}{\hat{\vec{q}}_{1}^{\;2}\hat{\vec{q}
}_{2}^{\;2}}|\bar{B}^{\prime}B\rangle\rightarrow{\hat{\mathcal{O}}^{-1}}
\frac{1}{\hat{\vec{q}}_{1}^{\;2}\hat{\vec{q}}_{2}^{\;2}}|\bar{B}^{\prime
}B\rangle\;, \label{kernel transformation}
\end{equation}
which does not change the discontinuity (\ref{discontinuity representation}).
In (\ref{kernel transformation}) $\hat{\mathcal{O}}$ is an arbitrary
nonsingular operator. The kernel $\hat{\mathcal{K}}$ in (\ref{kernel ff}) is
related with the one defined in Ref.~\cite{FF98} by such transformation with
$\hat{\mathcal{O}}=(\hat{\vec{q}}_{1}^{\;2}\hat{\vec{q}}_{2}^{\;2})^{1/2}$.
The reason for this choice is that in the leading order the kernel which is
conformal invariant and simply related to the dipole kernel is not the kernel
defined in Ref.~\cite{FF98}, but just the kernel $\hat{\mathcal{K}}$ in
(\ref{kernel ff})~\cite{Lipatov:1985uk,Bartels:2004ef}. Note that after the
choice of the operator $\hat{\mathcal{O}}$ in the leading order, additional
transformations with $\hat{\mathcal{O}}=1-\hat{O}$, where $\hat{O}\sim g^{2}$,
are still possible. At the NLO after such transformation we get
\begin{equation}
\hat{\mathcal{K}}\rightarrow\hat{\mathcal{K}}-[\hat{\mathcal{K}}^{(B)},\hat
{O}]~, \label{transformation at NLO}
\end{equation}
where $\hat{\mathcal{K}}^{(B)}$ is the leading order kernel.

In the NLO the BFKL kernel contains both gluon and quark contributions.
Refs.~\cite{Fadin:2006ha, Fadin:2007ee} were devoted to the quark
contribution. In Ref.~\cite{Fadin:2006ha} the \textquotedblleft
non-Abelian\textquotedblright\ part of the quark contribution~\cite{FFP99} was
transformed to the coordinate representation. The transformation was carried
out in the most general way: at arbitrary $D$ and for the case of arbitrary
impact factors. Generally speaking, the BFKL and dipole kernels are not
equivalent. But in case of scattering of colorless objects, besides the
freedom in the definition of the kernel discussed above, there is an additional
freedom related to the \textquotedblleft gauge invariance\textquotedblright
\ (vanishing at zero Reggeized gluon momenta) of the impact
factors~\cite{Lipatov:1985uk,Bartels:2004ef}. In this case the kernel
$\langle\vec{r}_{1}\vec{r}_{2}|\hat{\mathcal{K}}|\vec{r}_{1}^{\;\prime}\vec
{r}_{2}^{\;\prime}\rangle$ can be written in the dipole form (see below).
Owing to the possibility of the operator transformations
(\ref{transformation at NLO}), the dipole form is not unique. It was shown in
Ref.~\cite{Fadin:2006ha} that after the transformation
(\ref{transformation at NLO}) with a suitable operator $\hat{O}$ the
\textquotedblleft non-Abelian\textquotedblright\ part of the quark
contribution~\cite{FFP99} to the BFKL kernel, transferred to the dipole form,
agrees with the result obtained recently in Ref.~\cite{Balitsky:2006wa} by 
direct calculation of the quark contribution to the BK kernel in the
coordinate representation.

In Ref.~\cite{Fadin:2007ee} the \textquotedblleft Abelian\textquotedblright
part of the quark contribution~\cite{FFP99} was considered and its dipole
form was found. Since the \textquotedblleft Abelian\textquotedblright\ part is
known only in the limit $\epsilon\rightarrow0$, the approach adopted in
Ref.~\cite{Fadin:2007ee} was analogous to the one used in Section 5 of
Ref.~\cite{Fadin:2006ha}, i.e. the starting point was the renormalized BFKL
kernel, which was then simplified by the cancellation of the infrared
singularities between virtual and real contributions and presented at $D=4$.
After that its dipole form was found, which turned out to coincide with the
corresponding part of the quark contribution to the BK kernel calculated in
Ref.~\cite{Balitsky:2006wa} and to be evidently conformal invariant.

Here we consider the gluon part of the kernel (in the following sometimes for
brevity we will call this part simply ``kernel'') and find its dipole form. Evidently
this is the main and the most important part. Since the gluon part is known
only in the limit $\epsilon\rightarrow0$, we adopt the same strategy as in
Section 5 of Ref.~\cite{Fadin:2006ha} and in Ref.~\cite{Fadin:2007ee}.

\section{Decomposition of the gluon contribution}

\label{sec:decomposition}

The real part of the gluon contribution to the BFKL kernel in the colour
singlet channel is written~\cite{FF05} as
\begin{equation}
\hat{\mathcal{K}}_{r}=2\hat{\mathcal{K}}_{r}^{({8})}+\frac{1}{2}
\hat{\mathcal{K}}_{GG}^{({s})}~, \label{decomposition}
\end{equation}
where $\hat{\mathcal{K}}_{r}^{({8})}$ is the real part of the colour octet
kernel, concerned with the gluon Reggeization, and $\hat{\mathcal{K}}
_{GG}^{({s})}$ is the so called \textquotedblleft symmetric\textquotedblright
\ part of the two-gluon contribution to the non-forward BFKL kernel. The first
of them was calculated in Ref.~\cite{{FG00}}, and the second in
Ref.~\cite{FF05}. It is convenient to consider them separately. The use of the
decomposition (\ref{decomposition}) is due to the fact that all infrared
singularities turn out to be located in $\hat{\mathcal{K}}_{r}^{({8})}$,
whereas the \textquotedblleft symmetric\textquotedblright\ part $\hat
{\mathcal{K}}_{GG}^{({s})}$ is infrared finite. Moreover, in the momentum
representation $\hat{\mathcal{K}}_{r}^{({8})}$ looks much simpler than
$\hat{\mathcal{K}}_{GG}^{({s})}$. The origin of the relative simplicity of
$\hat{\mathcal{K}}_{r}^{({8})}$ is that only planar diagrams do contribute to
it. The \textquotedblleft symmetric\textquotedblright\ part includes
contributions from both planar and non-planar diagrams, the latter being the
most complicated.

At fixed non-zero $\vec k^{2}$, when the term $({\vec
k^{\;2}}/{\mu^{2} })^{\epsilon}$ can be expanded in powers of
$\epsilon$, the piece $\langle
\vec{q}_{1},\vec{q}_{2}|\hat{\mathcal{K}}^{({8})}_{r}|
\vec{q}^{\;\prime} _{1},\vec{q}^{\;\prime}_{2}\rangle$ is finite
at $\epsilon=0$. But this part is singular at $\vec k^{\;2}=0$,
therefore the region of $\vec k^{\;2}$ so small that
$\epsilon|\ln({\vec k^{\;2}}/{\mu^{2}})|\sim1$ and the expansion
of $({\vec k^{\;2}}/{\mu^{2}})^{\epsilon}$ cannot be done, is
important. Moreover, terms of order $\epsilon$ must be taken into
account in the coefficient of the expression divergent at $\vec
k^{\;2}=0$, in order to take all contributions non-vanishing in
the limit $\epsilon\rightarrow0$ after the integration over $\vec
k$.

The total BFKL kernel in the singlet case must be free from singularities.
Therefore, the infrared singularities of $2\hat{\mathcal{K}}_{r}^{({8})}$ must
cancel the singularities of the \textquotedblleft virtual\textquotedblright
\ contribution $\hat{\omega}_{1}+\hat{\omega}_{2}$ in
(\ref{operator of the BFKL kernel}). In the following we join these parts. The
experience of the transformation of the \textquotedblleft
non-Abelian\textquotedblright\ part of the quark
contribution~\cite{Fadin:2006ha} teaches us that the kernel in the coordinate
representation can be simplified by the operator transformation
(\ref{transformation at NLO}) with an appropriate $\hat{O}$. Here we will use
the transformed kernel with
\begin{equation}
\hat{O}=-\frac{11\alpha_{s}(\mu)N_{c}}{24\pi}\ln\left(  \hat{\vec{q}}
_{1}^{\;2}\hat{\vec{q}}_{2}^{\;2}\right)  ~, \label{choice of O}
\end{equation}
and call
\begin{equation}
\hat{\mathcal{K}}_{{p}}=\hat{\omega}_{1}+\hat{\omega}_{2}+2\hat{\mathcal{K}
}_{r}^{({8})}+\frac{11\alpha_{s}(\mu)N_{c}}{24\pi}\biggl[\hat{\mathcal{K}
}^{(B)},\ln\left(  \hat{\vec{q}}_{1}^{\;2}\hat{\vec{q}}_{2}^{\;2}\right)
\biggr] \label{planar kernel}
\end{equation}
\textquotedblleft planar\textquotedblright\ part. We put also $\hat
{\mathcal{K}}_{GG}^{({s})}=2\hat{\mathcal{K}}_{s}$, call $\hat{\mathcal{K}
}_{s}$ \textquotedblleft symmetric\textquotedblright\ part of the kernel and
write the kernel as
\begin{equation}
\hat{\mathcal{K}}=\hat{\mathcal{K}}_{{p}}+\hat{\mathcal{K}}_{{s}}~.
\label{new decomposition}
\end{equation}
The \textquotedblleft symmetric\textquotedblright\ part
$\langle\vec{q}
_{1},\vec{q}_{2}|\hat{\mathcal{K}}_{s}|\vec{q}_{1}^{\;\prime},\vec{q}
_{2}^{\;\prime}\rangle$ is finite in the limit $\epsilon=0$.
Moreover, it does not give terms divergent in $\epsilon=0$ by
action of the kernel, since it has no non-integrable singularities
in the limit $\epsilon=0$. Hence, we can consider the
\textquotedblleft symmetric\textquotedblright\ part in the
physical space-time dimension $D=4$ from the beginning.

\section{The NLO gluon trajectory}

\label{sec:trajectory}

We use the representation of the gluon trajectory in the form of integral
in the transverse momentum plane~\cite{Fadin:1995xg}. Taking into account that
the bare coupling $g$ in pure gluodynamics is connected with the renormalized
coupling $g_{\mu}$ in the ${\overline{\mbox{MS}}}$ scheme through the relation
\begin{equation}
g=g_{\mu}\mu^{-\mbox{\normalsize $\epsilon$}}\left[  1+\frac{11}{3}\frac
{\bar{g}_{\mu}^{2}}{2\epsilon}\right]  ~,\;\;\;\;\bar{g}_{\mu}^{2}
=\frac{g_{\mu}^{2}N_{c}\Gamma(1-\epsilon)}{(4\pi)^{2+{\epsilon}}}~,
\label{coupling renormalization}
\end{equation}
where $\Gamma(x)$ is the Euler gamma-function, we have
\begin{equation}
\omega(-\vec{q}_{i}^{\;2})=-\frac{\bar{g}_{\mu}^{2}\;\vec{q}_{i}^{\;2}}
{\pi^{1+\epsilon}\Gamma(1-\epsilon)}\int\frac{d^{2+2\epsilon}k\;\mu
^{-2\epsilon}}{\vec{k}^{\;2}(\vec{k}-\vec{q}_{i})^{2}}\Biggl(1+\bar{g}_{\mu
}^{2}\Biggl[\frac{11}{3\epsilon}-f(\vec{k},\vec{k}-\vec{q}_{i})+f(\vec
{k},0)+f(0,\vec{k}-\vec{q}_{i})\Biggr]\Biggr)~.
\label{omega as double integral}
\end{equation}
Here $\psi(x)=\Gamma^{\prime}(x)/\Gamma(x)$ and
\begin{equation}
f(\vec{k}_{1},\vec{k}_{2})=\frac{\vec{k}_{12}^{\;2}}{\pi^{1+\epsilon}
\Gamma(1-\epsilon)}\int\frac{d^{2+2\epsilon}q\;\mu^{-2\epsilon}}{(\vec{k}
_{1}-\vec{q})^{2}(\vec{k}_{2}-\vec{q})^{2}}\Biggl(\ln\left(  \frac{\vec
{k}_{12}^{\;2}}{\vec{q}^{\;2}}\right)  +a_{D}\Biggr)~, \label{f omega}
\end{equation}
where
\begin{equation}
a_{D}=-2\psi(D-3)-\psi\left(  3-\frac{D}{2}\right)  +2\psi\left(  \frac{D}
{2}-2\right)  +\psi(1)+\frac{2}{(D-3)(D-4)}+\frac{D-2}{4(D-1)(D-3)}~.
\label{a omega}
\end{equation}
The expressions (\ref{f omega}) and (\ref{a omega}) are exact in $\epsilon$. We need
to keep the terms of zero order in $\epsilon$ in the trajectory
(\ref{omega as double integral}), that means the terms of order $\epsilon$ in
$f(\vec{k}_{1},\vec{k}_{2})$ and of order $\epsilon^{2}$ in $a_{D}$. With the
required accuracy
\begin{equation}
a_{D}=-\frac{1}{\epsilon}-\frac{11}{6}+\epsilon\left(  \frac{67}{18}
-\zeta(2)\right)  -\epsilon^{2}\left(  \frac{202}{27}-7\zeta(3)\right)  ~,
\label{a omega at D near 4}
\end{equation}
where $\zeta(n)$ is the Riemann zeta-function. The integral
entering $f(\vec{k},0)=f(0,\vec{k})$ is known at arbitrary $D$
(see for instance Eq.~(B.16) in the second of Refs.~\cite{FG00}):
\begin{equation}
\frac{\vec{k}^{\;2}}{\pi^{1+\epsilon}\Gamma(1-\epsilon)}\int\frac
{d^{2+2\epsilon}q\;\mu^{-2\epsilon}}{\vec{q}^{\;2}(\vec{k}-\vec{q})^{2}}
\ln\left(  \frac{\vec{k}^{\;2}}{\vec{q}^{\;2}}\right)  \!=\!\frac{\Gamma
^{2}(\epsilon)}{\Gamma(2\epsilon)}\left(  \frac{\vec{k}^{\;2}}{\mu^{2}
}\right)  ^{\epsilon}\Biggl(\frac{1}{2\epsilon}-\psi(1)+\psi(1-\epsilon
)-\psi(1+\epsilon)+\psi(1+2\epsilon)\Biggr). \label{int1 for f}
\end{equation}
The integral for $f(\vec{k}_{1},\vec{k}_{2})$ was calculated with the required
accuracy in Ref.~\cite{Fadin:1999de} (see Eq.~(A.13)):
\[
\frac{\vec{k}_{12}^{\;2}}{\pi^{1+\epsilon}\Gamma(1-\epsilon)}\int
\frac{d^{2+2\epsilon}q\;\mu^{-2\epsilon}}{(\vec{k}_{1}-\vec{q})^{2}(\vec
{k}_{2}-\vec{q})^{2}}\ln\left(  \frac{\vec{k}_{12}^{\;2}}{\vec{q}^{\;2}
}\right)  =\frac{\Gamma^{2}(\epsilon)}{2\epsilon\Gamma(2\epsilon)}\left(
2\left(  \frac{\vec{k}_{12}^{\;2}}{\mu^{2}}\right)  ^{\epsilon}-\left(
\frac{\vec{k}_{1}^{\;2}}{\mu^{2}}\right)  ^{\epsilon}-\left(  \frac{\vec
{k}_{2}^{\;2}}{\mu^{2}}\right)  ^{\epsilon}\right)
\]
\begin{equation}
+\ln\left(  \frac{\vec{k}_{12}^{\;2}}{\vec{k}_{1}^{\;2}}\right)  \ln\left(
\frac{\vec{k}_{12}^{\;2}}{\vec{k}_{2}^{\;2}}\right)  -8\epsilon\zeta(3)~.
\label{int2 for f}
\end{equation}
We get then
\begin{equation}
\omega(-\vec{q}_{i}^{\;2})=-\frac{\bar{g}_{\mu}^{2}\;\vec{q}_{i}^{\;2}}
{\pi^{1+\epsilon}\Gamma(1-\epsilon)}\int\frac{d^{2+2\epsilon}k\;\mu
^{-2\epsilon}}{\vec{k}^{\;2}(\vec{k}-\vec{q}_{i})^{2}}\left(  1+\bar{g}_{\mu
}^{2}f_{\omega}(\vec{k},\vec{k}-\vec{q}_{i})\right)  ~,
\label{omega as integral}
\end{equation}
where, with the required accuracy,
\[
f_{\omega}(\vec{k}_{1},\vec{k}_{2})=\frac{11}{3\epsilon}-f(\vec{k}_{1},\vec
{k}_{2})+f(\vec{k}_{1},0)+f(0,\vec{k}_{2})=\frac{11}{3\epsilon}+\left[
\frac{11}{3\epsilon}-\frac{67}{9}+2\zeta(2)\right.
\]
\begin{equation}
\left.  +\epsilon\left(  \frac{404}{27}-\frac{11}{3}\zeta(2)-6\zeta(3)\right)
\right]  \left[  \left(  \frac{\vec{k}_{12}^{\;2}}{\mu^{2}}\right)
^{\epsilon}-\left(  \frac{\vec{k}_{1}^{\;2}}{\mu^{2}}\right)  ^{\epsilon
}-\left(  \frac{\vec{k}_{2}^{\;2}}{\mu^{2}}\right)  ^{\epsilon}\right]
-\ln\left(  \frac{\vec{k}_{12}^{\;2}}{\vec{k}_{1}^{\;2}}\right)  \ln\left(
\frac{\vec{k}_{12}^{\;2}}{\vec{k}_{2}^{\;2}}\right)  ~.
\label{integrand for omega}
\end{equation}
The representation (\ref{omega as integral})-(\ref{integrand for
omega}) is extremely convenient,  since it permits to get easily
the known expression for the trajectory in the limit
$\epsilon\rightarrow0$~ \cite{trajectory}.  But its main advantage
is that it gives the possibility to perform explicitly the
cancellation of the infrared singularities and to write the kernel
at the physical space-time dimension $D=4$, as it will be shown in
Section \ref{sec:cancellation}.

\section{The real part of the colour octet kernel}

\label{sec:octet}

The real part of the colour octet kernel (see Eq.~(68) of Ref.~\cite{FG00})
expressed through the renormalized coupling constant $\bar{g}_{\mu}$ has the
form
\[
\langle\vec{q}_{1},\vec{q}_{2}|\hat{\mathcal{K}}_{r}^{({8})}|\vec{q}
_{1}^{\;\prime},\vec{q}_{2}^{\;\prime}\rangle=\delta(\vec{q}-\vec{q}
^{\;\prime})\frac{\bar{g}_{\mu}^{2}\mu^{-2\epsilon}}{2\pi^{1+\epsilon}
\Gamma(1-\epsilon)\vec{q}_{1}^{\;2}\vec{q}_{2}^{\;2}}\left\{  \left(
\frac{\vec{q}_{1}^{\;2}\vec{q}_{2}^{\;\prime\;2}+\vec{q}_{1}^{\;\prime\;2}
\vec{q}_{2}^{\;2}}{\vec{k}^{\;2}}-\vec{q}^{\;2}\right)  \right.
\]
\[
\times\Biggl(1+\bar{g}_{\mu}^{2}\Biggl[\frac{11}{3\epsilon}+\left(  \frac
{\vec{k}^{\;2}}{\mu^{2}}\right)  ^{\epsilon}\left(  -\frac{11}{3\epsilon
}+\frac{67}{9}-2\zeta(2)+\epsilon\left(  -\frac{404}{27}+14\zeta(3)+\frac
{11}{3}\zeta(2)\right)  \right)  \Biggr]\Biggr)
\]
\[
+\bar{g}_{\mu}^{2}\Biggl[\vec{q}^{\,2}\left(  \frac{11}{3}\ln\left(
\frac{\vec{q}_{1}^{\;2}\vec{q}_{1}^{\;\prime\;2}}{\vec{q}^{\;2}\vec{k}^{\;2}
}\right)  +\frac{1}{2}\ln\left(  \frac{\vec{q}_{1}^{\;2}}{\vec{q}^{\;2}
}\right)  \ln\left(  \frac{\vec{q}_{2}^{\;2}}{\vec{q}^{\;2}}\right)  +\frac
{1}{2}\ln\left(  \frac{\vec{q}_{1}^{\;\prime\;2}}{\vec{q}^{\;2}}\right)
\ln\left(  \frac{\vec{q}_{2}^{\;\prime2}}{\vec{q}^{\;2}}\right)  \right.
\]
\[
\left.  +\frac{1}{2}\ln^{2}\left(  \frac{\vec{q}_{1}^{\;2}}{\vec{q}
_{1}^{\;\prime\;2}}\right)  \right)  -\frac{\vec{q}_{1}^{\;2}\vec{q}
_{2}^{\;\prime\;2}+\vec{q}_{2}^{\;2}\vec{q}_{1}^{\;\prime\;2}}{\vec{k}^{\;2}
}\ln^{2}\left(  \frac{\vec{q}_{1}^{\;2}}{\vec{q}_{1}^{\;\prime\;2}}\right)
+\frac{\vec{q}_{1}^{\;2}\vec{q}_{2}^{\;\prime\;2}-\vec{q}_{2}^{\;2}\vec{q}
_{1}^{\;\prime\;2}}{\vec{k}^{\;2}}\ln\left(  \frac{\vec{q}_{1}^{\;2}}{\vec
{q}_{1}^{\;\prime\;2}}\right)  \left(  \frac{11}{3}-\frac{1}{2}\ln\left(
\frac{\vec{q}_{1}^{\;2}\vec{q}_{1}^{\;\prime\;2}}{\vec{k}^{\;4}}\right)
\right)
\]
\[
\left.  +[\vec{q}^{\;2}(\vec{k}^{\;2}-\vec{q}_{1}^{\;2}-\vec{q}_{1}
^{\;\prime\;2})+2\vec{q}_{1}^{\;2}\vec{q}_{1}^{\;\prime\;2}-\vec{q}_{1}
^{\;2}\vec{q}_{2}^{\;\prime\;2}-\vec{q}_{2}^{\;2}\vec{q}_{1}^{\;\prime
\;2}+\frac{\vec{q}_{1}^{\;2}\vec{q}_{2}^{\;\prime\;2}-\vec{q}_{2}^{\;2}\vec
{q}_{1}^{\;\prime\;2}}{\vec{k}^{\;2}}(\vec{q}_{1}^{\;2}-\vec{q}_{1}
^{\;\prime\;2})]I(\vec{q}_{1}^{\;2},\vec{q}_{1}^{\;\prime\;2},\vec{k}
^{\;2})\Biggr]\right\}
\]
\begin{equation}
+\left(  \vec{q}_{1}\leftrightarrow\vec{q}_{2},\;\;\vec{q}_{1}^{\;\prime
}\leftrightarrow\vec{q}_{2}^{\;\prime}\right)  ~, \label{real octet}
\end{equation}
where
\begin{equation}
I(\vec{q}_{1}^{\;2},\vec{q}_{1}^{\;\prime\;2},\vec{k}^{\;2})=\int_{0}^{1}
\frac{dx}{\vec{q}_{1}^{\;2}(1-x)+\vec{q}_{1}^{\;\prime\;2}x-\vec{k}
^{\;2}x(1-x)}\ln\left(  \frac{\vec{q}_{1}^{\;2}(1-x)+\vec{q}_{1}^{\;\prime
\;2}x}{\vec{k}^{\;2}x(1-x)}\right)  ~. \label{integral I}
\end{equation}
Note that $I(a,b,c)$ is a totally symmetric function of the variables $a,\;b$
and $c$~\cite{Fadin:2002hz}, as it is obvious from the representation
\begin{equation}
I(a,b,c)=\int_{0}^{1}\int_{0}^{1}\int_{0}^{1}\frac{dx_{1}dx_{2}dx_{3}
\delta(1-x_{1}-x_{2}-x_{3})}{(ax_{1}+bx_{2}+cx_{3})(x_{1}x_{2}+x_{1}
x_{3}+x_{2}x_{3})}~.
\end{equation}
We will use also the representation
\begin{equation}
I(a,b,c)=\int_{0}^{1}dx\int_{0}^{1}{dz}\;\frac{1}{cx(1-x)z+(b(1-x)+ax)(1-z)}~.
\end{equation}
The leading singularity in (\ref{real octet}) is $1/\epsilon$. It turns again
into $\sim1/\epsilon^{2}$ after subsequent integrations of the kernel because
of the singular behaviour at $\vec{k}^{\;2}=0$. The additional singularity
arises from the region where $\epsilon|\ln\vec{k}^{\;2}/\mu^{2}|\sim1$.
For this reason, we have not expanded in $\epsilon$ the term $(\vec{k}^{\;2}/\mu
^{2})^{\epsilon}$. The terms $\sim\epsilon$ are taken into account in the
coefficient of the expression divergent at $\vec{k}^{\;2}=0$ in order to save
all non-vanishing contributions in the limit $\epsilon\rightarrow0$ after the integrations.

\section{Cancellation of the infrared singularities}

\label{sec:cancellation}

Let us introduce the cut-off $\lambda\rightarrow0$, making it tending to zero
after taking the limit $\epsilon\rightarrow0$, and divide the integration
region in the integral representation of the trajectory
(\ref{omega as integral}) into three domains. In two of them either $\vec
{k}^{\;2}\leq\lambda^{2}$, or $(\vec{k}-\vec{q}_{i})^{2}\leq\lambda^{2}$, and
in the third one both $\vec{k}^{\;2}>\lambda^{2}$ and $(\vec{k}-\vec{q}
_{i})^{2}>\lambda^{2}$. Then in the third domain we can take the limit
$\epsilon=0$ in (\ref{integrand for omega}) and put $f_{\omega}(\vec{k}
_{1},\vec{k}_{2})=f_{\omega}^{(0)}(\vec{k}_{1},\vec{k}_{2})$, where
\begin{equation}
f_{\omega}^{(0)}(\vec{k}_{1},\vec{k}_{2})=\frac{67}{9}-2\zeta(2)-\frac{11}
{3}\ln\left(  \frac{\vec{k}_{1}^{\;2}\vec{k}_{2}^{\;2}}{\mu^{2}\vec{k}
_{12}^{\;2}}\right)  -\ln\left(  \frac{\vec{k}_{12}^{\;2}}{\vec{k}_{1}^{\;2}
}\right)  \ln\left(  \frac{\vec{k}_{12}^{\;2}}{\vec{k}_{2}^{\;2}}\right)  ~.
\label{integrand for omega at large k}
\end{equation}
In the first domain we have
\begin{equation}
f_{\omega}(\vec{k},\vec{k}-\vec{q}_{i})=\frac{11}{3\epsilon}-\left(
\frac{\vec{k}^{\;2}}{\mu^{2}}\right)  ^{\epsilon}\Biggl[\frac{11}{3\epsilon
}-\frac{67}{9}+2\zeta(2)+\epsilon\left(  \frac{404}{27}-\frac{11}{3}
\zeta(2)-6\zeta(3)\right)  \Biggr]~, \label{integrand for omega at small k}
\end{equation}
and in the second one we have the same expression with the substitution
$\vec{k}^{\;2}\rightarrow(\vec{k}-\vec{q}_{i})^{2}$. Writing the real part of
the colour octet kernel as
\begin{equation}
\langle\vec{q}_{1},\vec{q}_{2}|\hat{\mathcal{K}}_{r}^{({8})}|\vec{q}
_{1}^{\;\prime},\vec{q}_{2}^{\;\prime}\rangle=\langle\vec{q}_{1},\vec{q}
_{2}|\hat{\mathcal{K}}_{r}^{({8})}|\vec{q}_{1}^{\;\prime},\vec{q}
_{2}^{\;\prime}\rangle\theta(\lambda^{2}-\vec{k}^{\;2})+\langle\vec{q}
_{1},\vec{q}_{2}|\hat{\mathcal{K}}_{r}^{({8})}|\vec{q}_{1}^{\;\prime},\vec
{q}_{2}^{\;\prime}\rangle\theta(\vec{k}^{\;2}-\lambda^{2})~,
\label{decomposition of K 8r}
\end{equation}
and comparing (\ref{real octet}) with (\ref{integrand for omega at small k})
we see that in the \textquotedblleft planar\textquotedblright\ kernel
(\ref{planar kernel}) the first term in the R.H.S. of
(\ref{decomposition of K 8r}) cancels almost completely the contributions of
the regions $\vec{k}^{\;2}\leq\lambda^{2}$ and $(\vec{k}-\vec{q}_{i})^{2}
\leq\lambda^{2}$ in the trajectories $\omega(-\vec{q}_{i}^{\;2})$. The only
piece which remains uncancelled in each of the trajectories for $\epsilon
\rightarrow0$ is
\begin{equation}
\frac{\bar{g}_{\mu}^{4}\;}{\pi^{1+\epsilon}\Gamma(1-\epsilon)}\int
\frac{d^{2+2\epsilon}k\;\mu^{-2\epsilon}}{\vec{k}^{\;2}}16\epsilon
\zeta(3)\left(  \frac{\vec{k}^{\;2}}{\mu^{2}}\right)  ^{\epsilon}
\theta(\lambda^{2}-\vec{k}^{\;2})=\frac{\alpha_{s}^{2}(\mu)N_{c}^{2}}{2\pi
^{2}}\zeta(3). \label{uncancelled piece of omega}
\end{equation}
On account of this cancellation and using the equality
\begin{equation}
\int\frac{d^{2}k\;}{4\pi}\frac{\vec{q}^{\;2}}{\vec{k}^{\;2}(\vec{k}-\vec
{q})^{2}}\ln\left(  \frac{\vec{k}^{\;2}}{\vec{q}^{\;2}}\right)  \ln\left(
\frac{(\vec{k}-\vec{q})^{2}}{\vec{q}^{\;2}}\right)  =\zeta(3)~,
\label{representation for zeta}
\end{equation}
we can put
\[
\langle\vec{q}_{1},\vec{q}_{2}|\hat{\omega}_{1}+\hat{\omega}_{2}
+2\hat{\mathcal{K}}_{r}^{({8})}|\vec{q}_{1}^{\;\prime},\vec{q}_{2}^{\;\prime
}\rangle=-\delta(\vec{q}_{11^{\prime}})\delta(\vec{q}_{22^{\prime}}
)\frac{\alpha_{s}(\mu)N_{c}}{4\pi^{2}}
\]
\[
\times\Biggl(\int d^{2}k\;\left(  \frac{2}{\vec{k}^{\;2}}+2\frac{\vec{k}
(\vec{q}_{1}-\vec{k})}{\vec{k}^{\;2}(\vec{q}_{1}-\vec{k})^2}+\frac{\alpha
_{s}(\mu)N_{c}}{\pi}\left(  V(\vec{k})+V(\vec{k},\vec{k}-\vec{q}_{1})\right)
\right)  -3\alpha_{s}(\mu)N_{c}\zeta(3)\Biggr)
\]
\[
+\delta(\vec{q}-\vec{q}^{\;\prime})\frac{\alpha_{s}(\mu)N_{c}}{4\pi^{2}\vec
{q}_{1}^{\;2}\vec{q}_{2}^{\;2}}\left\{  \left(  \frac{\vec{q}_{1}^{\;2}\vec
{q}_{2}^{\;\prime\;2}+\vec{q}_{1}^{\;\prime\;2}\vec{q}_{2}^{\;2}}{\vec
{k}^{\;2}}-\vec{q}^{\;2}\right)  \Biggl(1+\frac{\alpha_{s}(\mu)N_{c}}{4\pi
}\Biggl[-\frac{11}{3}\ln\left(  \frac{\vec{k}^{\;2}}{\mu^{2}}\right)
+\frac{67}{9}-2\zeta(2)\Biggr]\Biggr)\right.
\]
\[
+\frac{\alpha_{s}(\mu)N_{c}}{4\pi}\Biggl[\vec{q}^{\,2}\left(  \frac{11}{3}
\ln\left(  \frac{\vec{q}_{1}^{\;2}\vec{q}_{1}^{\;\prime\;2}}{\vec{q}^{\;2}
\vec{k}^{\;2}}\right)  +\frac{1}{2}\ln\left(  \frac{\vec{q}_{1}^{\;2}}{\vec
{q}^{\;2}}\right)  \ln\left(  \frac{\vec{q}_{2}^{\;2}}{\vec{q}^{\;2}}\right)
+\frac{1}{2}\ln\left(  \frac{\vec{q}_{1}^{\;\prime\;2}}{\vec{q}^{\;2}}\right)
\ln\left(  \frac{\vec{q}_{2}^{\;\prime2}}{\vec{q}^{\;2}}\right)  \right.
\]
\[
\left.  +\frac{1}{2}\ln^{2}\left(  \frac{\vec{q}_{1}^{\;2}}{\vec{q}
_{1}^{\;\prime\;2}}\right)  \right)  -\frac{\vec{q}_{1}^{\;2}\vec{q}
_{2}^{\;\prime\;2}+\vec{q}_{2}^{\;2}\vec{q}_{1}^{\;\prime\;2}}{\vec{k}^{\;2}
}\ln^{2}\left(  \frac{\vec{q}_{1}^{\;2}}{\vec{q}_{1}^{\;\prime\;2}}\right)
+\frac{\vec{q}_{1}^{\;2}\vec{q}_{2}^{\;\prime\;2}-\vec{q}_{2}^{\;2}\vec{q}
_{1}^{\;\prime\;2}}{\vec{k}^{\;2}}\ln\left(  \frac{\vec{q}_{1}^{\;2}}{\vec
{q}_{1}^{\;\prime\;2}}\right)  \left(  \frac{11}{3}-\frac{1}{2}\ln\left(
\frac{\vec{q}_{1}^{\;2}\vec{q}_{1}^{\;\prime\;2}}{\vec{k}^{\;4}}\right)
\right)
\]
\[
\left.  +\biggl[\vec{q}^{\;2}(\vec{k}^{\;2}-\vec{q}_{1}^{\;2}-\vec{q}
_{1}^{\;\prime\;2})+2\vec{q}_{1}^{\;2}\vec{q}_{1}^{\;\prime\;2}-\vec{q}
_{1}^{\;2}\vec{q}_{2}^{\;\prime\;2}-\vec{q}_{2}^{\;2}\vec{q}_{1}^{\;\prime
\;2}+\frac{\vec{q}_{1}^{\;2}\vec{q}_{2}^{\;\prime\;2}-\vec{q}_{2}^{\;2}\vec
{q}_{1}^{\;\prime\;2}}{\vec{k}^{\;2}}(\vec{q}_{1}^{\;2}-\vec{q}_{1}
^{\;\prime\;2})\biggr]I(\vec{q}_{1}^{\;2},\vec{q}_{1}^{\;\prime\;2},\vec
{k}^{\;2})\Biggr]\!\!\right\}
\]
\begin{equation}
+\left(  \vec{q}_{1}\leftrightarrow\vec{q}_{2},\;\;\vec{q}_{1}^{\;\prime
}\leftrightarrow\vec{q}_{2}^{\;\prime}\right)  ~, \label{semiplanar
kernel at D=4}
\end{equation}
where
\begin{equation}
V(\vec{k})=\frac{1}{2\vec{k}^{\;2}}\left(  \frac{67}{9}-2\zeta(2)-\frac{11}
{3}\ln\left(  \frac{\vec{k}^{\;2}}{\mu^{2}}\right)  \right)  ~,
\label{function V1}
\end{equation}
\begin{eqnarray}
V(\vec{k},\vec{q})&=&\frac{\vec{k}\vec{q}}{2\vec{k}^{\;2}\vec{q}^{\;2}}\left(
\frac{11}{3}\ln\left(  \frac{\vec{k}^{\;2}\vec{q}^{\;2}}{\mu^{2}(\vec{k}
-\vec{q})^{2}}\right)  -\frac{67}{9}+2\zeta(2)\right)  -\frac{11}{12\vec
{k}^{\;2}}\ln\left(  \frac{\vec{q}^{\;2}}{(\vec{k}-\vec{q})^{2}}\right)
\label{function V2} \\
&-&\frac{11}{12\vec{q}^{\;2}}\ln\left(  \frac{\vec{k}^{\;2}}{(\vec{k}-\vec
{q})^{2}}\right)  ~. \nonumber
\end{eqnarray}
Thus we have obtained this piece of the kernel at $D=4$. Of course, the
infrared singularities in (\ref{semiplanar kernel at D=4}) must be regularized
either by limitations on the integration regions as discussed above or in an
equivalent way.

\section{The ``symmetric'' part of the kernel}

\label{sec:symmetric part}

The \textquotedblleft symmetric\textquotedblright\ part of the kernel was
found in the momentum representation in Ref.~\cite{FF05}. It contains neither
ultraviolet nor infrared singularities and therefore it does not require
regularization and renormalization. For this reason one can use from the
beginning physical space-time dimension $D=4$ and renormalized coupling
constant $\alpha_{s}(\mu)$. Nevertheless, in the momentum representation the
\textquotedblleft symmetric\textquotedblright\ part is the most complicated
piece of the gluon contribution to the kernel. In this respect the
\textquotedblleft symmetric\textquotedblright\ part is analogous to the
\textquotedblleft Abelian\textquotedblright\ part of the quark contribution
which was considered in Ref.~\cite{Fadin:2007ee} and turned out to be
surprisingly simple in the coordinate representation. But contrary to the
\textquotedblleft Abelian\textquotedblright\ part, the \textquotedblleft
symmetric\textquotedblright\ part contains the subtraction term which makes
its transformation more complicated and its form in the coordinate
representation considerably intricate.

As well as for the \textquotedblleft Abelian\textquotedblright\
part it is better to start from the expressions for the
\textquotedblleft symmetric\textquotedblright\ part before
integration over the momenta of the produced particles. The
starting point is Eq.~(3.43) of Ref.~\cite{FF05}, where a
symmetrization operator appears. The use of this symmetrization
operator is inconvenient, because our definition of the kernel
(\ref{kernel ff}) differs from the definition of Ref.~\cite{FF05}.
Instead, we reconstruct the explicit symmetry by restoring the
term represented by the last line in Eq.~(3.40) of
Ref.~\cite{FF05}, which was omitted in Eq.~(3.43). As a result, we
decompose the \textquotedblleft symmetric\textquotedblright\ part
into two pieces:
\begin{equation}
\hat{\mathcal{K}}_{s}=\hat{\mathcal{K}}_{s1}+\hat{\mathcal{K}}_{s2}~,
\label{decomposition of K s}
\end{equation}
where
\begin{equation}
\langle\vec{q}_{1},\vec{q}_{2}|\hat{\mathcal{K}}_{s1}|\vec{q}_{1}^{\;\prime
},\vec{q}_{2}^{\;\prime}\rangle=\delta(\vec{q}-\vec{q}^{\;\prime})\frac
{1}{\vec{q}_{1}^{\;2}\vec{q}_{2}^{\;2}}\frac{\alpha_{s}^{2}(\mu)N_{c}^{2}
}{2\pi^{3}}\int_{0}^{1}{dx}\int\frac{d^{2}k_{1}}{2\pi}\left(  \frac
{F_{s}(k_{1},k_{2})}{x(1-x)}\right)  _{+} \label{Ks1}
\end{equation}
and
\begin{equation}
\langle\vec{q}_{1},\vec{q}_{2}|\hat{\mathcal{K}}_{s2}|\vec{q}_{1}^{\;\prime
},\vec{q}_{2}^{\;\prime}\rangle=-\delta(\vec{q}-\vec{q}^{\;\prime})\int
\frac{d^{2}k_{1}}{4\vec{q}_{1}^{\;2}\vec{q}_{2}^{\;2}}\frac{\mathcal{K}
_{r}^{B}(\vec{q}_{1},\vec{q}_{1}-\vec{k}_{1};\vec{q})\mathcal{K}_{r}^{B}
(\vec{q}_{1}-\vec{k}_{1},\vec{q}_{1}^{\;\prime};\vec{q})}{(\vec{q}_{1}-\vec
{k}_{1})^{2}(\vec{q}_{2}+\vec{k}_{1})^{2}}\ln\left(  \frac{\vec{k}_{2}^{\;2}
}{\vec{k}_{1}^{\;2}}\right)  ~. \label{K s2}
\end{equation}
Here $\vec{k}_{1}+\vec{k}_{2}=\vec{k}=\vec{q}_{11^{\prime}}$, the subscript
$_{+}$ means
\begin{equation}
\left(  \frac{f(x)}{x(1-x)}\right)  _{+}\equiv\frac{1}{x}[f(x)-f(0)]+\frac
{1}{(1-x)}[f(x)-f(1)]~, \label{integral +}
\end{equation}
the function $F_{s}(k_{1},k_{2})$ is defined in Eqs.~(3.44), (3.45) and (4.1)
of Ref.~\cite{FF05}, and $\mathcal{K}_{r}^{B}$ is the real part of the leading
order kernel,
\begin{equation}
\mathcal{K}_{r}^{B}(\vec{q}_{1},\vec{q}_{1}^{\;\prime};\vec{q})=\frac
{\alpha_{s}(\mu)N_{c}}{2\pi^{2}}\left(  \frac{\vec{q}_{1}^{\;2}\vec{q}
_{2}^{\;\prime\;2}+\vec{q}_{1}^{\;\prime\;2}\vec{q}_{2}^{\;2}}{\vec{k}^{\;2}
}-\vec{q}^{\;2}\right)  ~. \label{real born}
\end{equation}

\section{Dipole form of the kernel}

\label{sec:Dipole form}

With our normalizations, the kernel $\hat{\mathcal{K}}$ in the coordinate
representation is given by
\begin{equation}
\langle\vec{r}_{1}\vec{r}_{2}|\hat{\mathcal{K}}|\vec{r}_{1}^{\;\prime}\vec
{r}_{2}^{\;\prime}\rangle=\int\frac{d^{2}q_{1}}{2\pi}\frac{d^{2}q_{2}}{2\pi
}\frac{d^{2}q_{1}^{\prime}}{2\pi}\frac{d^{2}q_{2}^{\prime}}{2\pi}\langle
\vec{q}_{1},\vec{q}_{2}|\hat{\mathcal{K}}_{p}+\hat{\mathcal{K}}_{s}|\vec
{q}_{1}^{\;\prime},\vec{q}_{2}^{\;\prime}\rangle e^{i[\vec{q}_{1}\vec{r}
_{1}+\vec{q}_{2}\vec{r}_{2}-\vec{q}_{1}^{\;\prime}\vec{r}_{1}^{\;\prime}
-\vec{q}_{2}^{\;\prime}\vec{r}_{2}^{\;\prime}]}\;. \label{transformation to r}
\end{equation}
The real part of the kernel $\langle\vec{q}_{1},\vec{q}_{2}|\hat{\mathcal{K}
}_{r}|\vec{q}_{1}^{\;\prime},\vec{q}_{2}^{\;\prime}\rangle$ vanishes when any
of the $\vec{q}_{i}$'s or $\vec{q}_{i}^{\;\prime}$'s tends to zero. This
property, together with the \textquotedblleft gauge
invariance\textquotedblright\ property of the impact factors of colorless
\textquotedblleft projectiles\textquotedblright\ ${\Phi}_{A^{\prime}A}
(0,\vec{q})={\Phi}_{A^{\prime}A}(\vec{q},0)=0\;,$ permits us to change the
\textquotedblleft target\textquotedblright\ impact factors
in~(\ref{discontinuity representation}) so that they acquire the
\textquotedblleft dipole\textquotedblright\ property
\begin{equation}
\langle\vec{r},\vec{r}|({\hat{\vec{q}}_{1}^{\;2}\hat{\vec{q}}_{2}^{\;2}}
)^{-1}|\bar{B}^{\prime}B\rangle_{d}=0\;. \label{input}
\end{equation}
Further, adding the terms independent either of $\vec{r}_{1}$ or of $\vec
{r}_{2}$ (that is possible due to the \textquotedblleft gauge
invariance\textquotedblright\ ) one can change the kernel $\langle\vec{r}
_{1}\vec{r}_{2}|\hat{\mathcal{K}}|\vec{r}_{1}^{\;\prime}\vec{r}_{2}^{\;\prime
}\rangle$ so that it conserves the \textquotedblleft dipole\textquotedblright
\ property. Thereafter, terms proportional to $\delta(\vec{r}_{1}^{\;\prime}-
\vec{r}_{2}^{\;\prime})$ in $\langle\vec{r}_{1}\vec{r}_{2}|\hat{\mathcal{K}}|\vec{r}
_{1}^{\;\prime}\vec{r}_{2}^{\;\prime}\rangle$ can be omitted (see
Ref.~\cite{Fadin:2006ha} for details). We call the remaining part
$\hat{\mathcal{K}}_{d}$ ``the dipole form of the BFKL kernel''. In the LO,
discussed in detail in Ref.~\cite{Fadin:2006ha}, it coincides with the dipole
kernel:
\begin{equation}
\langle\vec{r}_{1}\vec{r}_{2}|\hat{\mathcal{K}}_{d}^{LO}|\vec{r}_{1}
^{\;\prime}\vec{r}_{2}^{\;\prime}\rangle=\frac{\alpha_{s}(\mu)N_{c}}{2\pi^{2}
}\int d\vec{\rho}\frac{\vec{r}_{12}{}^{2}}{\vec{r}_{1\rho}^{\,\,2}\vec
{r}_{2\rho}^{\,\,2}}\Biggl[\delta(\vec{r}_{11^{\prime}})\delta(\vec
{r}_{2^{\prime}\rho})+\delta(\vec{r}_{1^{\prime}\rho})\delta(\vec
{r}_{22^{\prime}})-\delta(\vec{r}_{11^{\prime}})\delta({r}_{22^{\prime}
})\Biggr]~. \label{LO dipole}
\end{equation}
Here $\vec{r}_{i\rho}=\vec{r}_{i}-$\ $\vec{\rho}$ and
$\vec{r}_{ij^{\prime}}=\vec{r}_{i}-\vec{r}_j^{\;\prime}$.

Note that the integrand in (\ref{LO dipole}) contains ultraviolet
singularities at $\vec{\rho}=\vec{r}_{1}$ and $\vec{\rho}=\vec{r}_{2}$ which
cancel in the sum of the contributions with account of the \textquotedblleft
dipole\textquotedblright\ property of the \textquotedblleft
target\textquotedblright\ impact factors. The coefficient of $\delta(\vec
{r}_{11^{\prime}})\delta({\vec{r}}_{22^{\prime}})$ is written in the integral
form in order to make the cancellation evident. The singularities do not
permit us to perform the integration in this coefficient.

In the NLO the dipole form can be written as
\[
\langle\vec{r}_{1}\vec{r}_{2}|\hat{\mathcal{K}}_{d}^{NLO}|\vec{r}
_{1}^{\;\prime}\vec{r}_{2}^{\;\prime}\rangle=\frac{\alpha_{s}^{2}(\mu
)N_{c}^{2}}{4\pi^{3}}\Biggl[\delta(\vec{r}_{11^{\prime}})\delta(\vec
{r}_{22^{\prime}})\int d\vec{\rho}\,g^{0}(\vec{r}_{1},\vec{r}_{2};\vec{\rho})
\]
\begin{equation}
+\delta(\vec{r}_{11^{\prime}})g(\vec{r}_{1},\vec{r}_{2};\vec{r}_{2}^{\;\prime
})+\delta(\vec{r}_{22^{\prime}})g(\vec{r}_{2},\vec{r}_{1};\vec{r}
_{1}^{\;\prime})+\frac{1}{\pi}g(\vec{r}_{1},\vec{r}_{2};\vec{r}_{1}^{\;\prime
},\vec{r}_{2}^{\;\prime})\Biggr]\; \label{kernel through g}\;,
\end{equation}
with the functions $g$ turning into zero  when their first two
arguments coincide.

The first three terms in the R.H.S. of (\ref{kernel through g}) contain ultraviolet
singularities which cancel in their sum, as well as in the LO, with account of
the \textquotedblleft dipole\textquotedblright\ property of the
\textquotedblleft target\textquotedblright\ impact factors. The coefficient of
$\delta(\vec{r}_{11^{\prime}})\delta({\vec{r}}_{22^{\prime}})$ is written in
the integral form in order to make the cancellation evident.

In the next Sections we will find the functions $g$.

\section{Transformation of the ``planar'' part}

\label{sec:planar}

Using (\ref{planar kernel}) and (\ref{semiplanar kernel at D=4}) and omitting
terms with $\delta(\vec{r}_{1^{\prime}2^{\prime}})$ in the coordinate space,
we reduce the NLO piece of the \textquotedblleft planar\textquotedblright
\ part to the form:
\[
\langle\vec{q}_{1},\vec{q}_{2}|\hat{\mathcal{K}}_{p}^{NLO}|\vec{q}
_{1}^{\;\prime},\vec{q}_{2}^{\;\prime}\rangle\rightarrow\frac{\alpha_{s}
^{2}(\mu)N_{c}^{2}}{4\pi^{3}}\Biggl[-\delta(\vec{q}_{11^{\prime}})\delta
(\vec{q}_{22^{\prime}})\Biggl(\int d\vec{k}\left(  V(\vec{k})+V(\vec{k}
,\vec{k}-\vec{q}_{1})\right)  -3\pi\zeta(3)\Biggr)
\]
\[
+\delta(\vec{q}-\vec{q}^{\;\prime})\left\{  V(\vec{k})+2V(\vec{k},\vec{q}
_{1})+\frac{(\vec{q}_{1}\vec{q}_{2})}{4\vec{q}_{1}^{\;2}\vec{q}_{2}^{\;2}
}\left[  \ln\left(  \frac{\vec{q}_{1}^{\;\prime\,2}}{\vec{q}^{\;2}}\right)
\ln\left(  \frac{\vec{q}_{2}^{\;\prime\,2}}{\vec{q}^{\;2}}\right)  +\ln
^{2}\left(  \frac{\vec{q}_{1}^{\;2}}{\vec{q}_{1}^{\;\prime\,2}}\right)
\right]  -\frac{1}{2\vec{k}^{2}}\ln^{2}\left(  \frac{\vec{q}_{1}^{\;2}}
{\vec{q}_{1}^{\;\prime\,2}}\right)  \right.
\]
\[
+\left[  \frac{(\vec{q}_{1}\vec{k})^{\;2}}{\vec{q}_{1}^{\;2}\vec{k}^{\;2}
}-1-\frac{(\vec{q}_{1}+\vec{k}\,)\vec{q}_{2}}{\vec{q}_{2}^{\;2}}+\left(
\frac{(\vec{k}\vec{q}_{2})}{\vec{k}^{\;2}}+\frac{(\vec{q}_{1}\vec
{q}_{2})}{\vec{q}_{1}^{\;2}}\right)  \frac{(\vec{q}_{1}\vec{k})}{\vec
{q}_{2}^{\;2}}\right]  I(\vec{q}_{1}^{\;2},\vec{q}_{1}^{\;\prime\,2},\vec
{k}^{\;2})
\]
\[
+\frac{(\vec{k}\vec{q}_{1})}{2\vec{k}^{\;2}\vec{q}_{1}^{\;2}}\left[  \ln
^{2}\left(  \frac{\vec{q}_{2}^{\;2}}{\vec{q}_{2}^{\;\prime\,2}}\right)
+\frac{1}{2}\ln\left(  \frac{\vec{q}_{2}^{\;2}}{\vec{q}_{2}^{\;\prime\,2}
}\right)  \ln\left(  \frac{\vec{q}_{2}^{\;2}\vec{q}_{2}^{\;\prime\,2}}{\vec
{k}^{\;4}}\right)  +\ln^{2}\left(  \frac{\vec{q}_{1}^{\;2}}{\vec{q}
_{1}^{\;\prime\,2}}\right)  -\frac{1}{2}\ln\left(  \frac{\vec{q}_{1}^{\;2}
}{\vec{q}_{1}^{\;\prime\,2}}\right)  \ln\left(  \frac{\vec{q}_{1}^{\;2}\vec
{q}_{1}^{\;\prime\,2}}{\vec{k}^{\;4}}\right)  \right]
\]
\begin{equation}
+\left.  \frac{1}{4\vec{q}_{1}^{\;2}}\left[  \ln\left(  \frac{\vec{q}
_{1}^{\;\prime\,2}}{\vec{k}^{\;2}}\right)  \ln\left(  \frac{\vec{q}_{1}
^{\;2}\vec{q}_{2}^{\;\prime\,2}}{\vec{q}_{2}^{\;2}\vec{q}_{1}^{\;\prime\,2}
}\right)  +\ln\left(  \frac{\vec{q}_{2}^{\;2}}{\vec{q}^{\;2}}\right)
\ln\left(  \frac{\vec{q}_{1}^{\;\prime\,2}\vec{q}_{2}^{\;\prime\,2}}{\vec
{q}^{\;4}}\right)  \right]  \right\}  +\left(  \vec{q}_{1}\leftrightarrow
\vec{q}_{2},\;\;\vec{q}_{1}^{\;\prime}\leftrightarrow\vec{q}_{2}^{\;\prime
}\right)  \Biggr]. \label{planar part for integration}
\end{equation}
Contributions to the coefficient of
$\delta(\vec{r}_{11^{\prime}})\delta(\vec {r}_{22^{\prime}})$ in
the square brackets of (\ref{kernel through g}) come only from the
terms with $V(\vec{k})$ and $\zeta(3)$ in (\ref{planar part for
integration}). The terms with $V(\vec{k})$ contribute to this
coefficient as  the integral
\begin{equation}
\tilde{V}(\vec{r}_{12})={2}\int d\vec{k}\,V(\vec{k})\left(
e^{i\vec{k}\,\vec {r}_{12}}-1\right)\;,  ~ \label{tilte V}
\end{equation}
with ultraviolet divergence.  The divergence appears as a result
of the separation of the ultraviolet non-singular sum
$V(\vec{k})+V(\vec{k},\vec {k}-\vec{q}_{1})$ into two pieces. We
have to represent the contribution of the terms with $V(\vec{k})$
in the integral form in order to make evident the cancellation of
ultraviolet singularities of separate terms in (\ref{kernel
through g}), as it was already mentioned in
Section~\ref{sec:Dipole form}. We use the same trick as in Section
5 of Ref.~\cite{Fadin:2006ha} and obtain with the help of the
integrals (\ref{int k/k^2}) and (\ref{int k/k^2 ln k/k^2}) of
Appendix A
\[
\tilde{V}(\vec{r}_{12})=-\int\frac{d\vec{k}\,d\vec{q}\,d\vec{\rho}}
{2(2\pi)^{2}}\frac{(\vec{q}\,\,\vec{k})}{\vec{q}^{\,\,2}\,\vec{k}^{\,\,2}
}\left(  \frac{67}{9}-2\zeta(2)-\frac{11}{6}\ln\left(  \frac{\vec{k}^{\;2}
\vec{q}^{\,\,2}}{\mu^{4}}\right)  \right)  \left(  2e^{i[\vec{k}\,\,\vec
{r}_{1\rho}+\vec{q}\,\vec{r}_{2\rho}]}\right.
\]
\begin{equation}
\left.  -e^{i(\vec{k}+\vec{q})\vec{r}_{1\rho}}-e^{i(\vec{k}+\vec{q})\vec
{r}_{2\rho}}\right)  =-\frac{11}{12}\int d\vec{\rho}\,\left[  \frac{\vec
{r}_{12}^{\;2}}{\vec{r}_{1\rho}^{\;2}\vec{r}_{2\rho}^{\;2}}\ln\left(
\frac{\vec{r}_{1\rho}^{\;2}\vec{r}_{2\rho}^{\;2}}{r_{\mu}^{4}}\right)
+\left(  \frac{1}{\vec{r}_{2\rho}^{\,\,\,2}}-\frac{1}{\vec{r}_{1\rho
}^{\,\,\,2}}\right)  \ln\left(  \frac{\vec{r}_{2\rho}^{\,\,\,2}}{\vec
{r}_{1\rho}^{\,\,\,2}}\right)  \right]  ~, \label{int V(k)}
\end{equation}
where
\begin{equation}
\ln r_{\mu}^{2}=2\psi(1)-\ln\frac{\mu^{2}}{4}-\frac{3}{11}\left(  \frac{67}
{9}-2\zeta(2)\right)  .
\end{equation}
For uniformity we write also the contribution of the term with $\zeta(3)$ in
integral form via the representation
\begin{equation}
\zeta(3)=\frac{1}{4\pi}\int d\vec{\rho}\frac{\,\vec{r}_{12}^{\;2}}{\vec
{r}_{1\rho}^{\;2}\vec{r}_{2\rho}^{\;2}}\ln\left(  \frac{\vec{r}_{1\rho}^{\;2}
}{\vec{r}_{12}^{\;2}}\right)  \ln\left(  \frac{\vec{r}_{2\rho}^{\;2}}{\vec
{r}_{12}^{\;2}}\right)  . \label{representation for zeta in coordinate space}
\end{equation}
Using the subscript $_{p}$ to denote the contributions of the
\textquotedblleft planar\textquotedblright\ part we have
\begin{equation}
g_{p}^{0}(\vec{r}_{1},\vec{r}_{2};\vec\rho)=\frac{3}{2}\frac{\,\vec{r}_{12}^{\;2}
}{\vec{r}_{1\rho}^{\;2}\vec{r}_{2\rho}^{\;2}}\ln\left(  \frac{\vec{r}_{1\rho
}^{\;2}}{\vec{r}_{12}^{\;2}}\right)  \ln\left(  \frac{\vec{r}_{2\rho}^{\;2}
}{\vec{r}_{12}^{\;2}}\right)  -\frac{11}{12}\left[  \frac{\vec{r}_{12}^{\;2}
}{\vec{r}_{1\rho}^{\;2}\vec{r}_{2\rho}^{\;2}}\ln\left(  \frac{\vec{r}_{1\rho
}^{\;2}\vec{r}_{2\rho}^{\;2}}{r_{\mu}^{4}}\right)  +\left(  \frac{1}{\vec
{r}_{2\rho}^{\,\,\,2}}-\frac{1}{\vec{r}_{1\rho}^{\,\,\,2}}\right)  \ln\left(
\frac{\vec{r}_{2\rho}^{\,\,\,2}}{\vec{r}_{1\rho}^{\,\,\,2}}\right)  \right]
~.
\end{equation}
The contribution to
$g(\vec{r}_{1},\vec{r}_{2};\vec{r}_{2}^{\;\prime})$ comes from all
terms which do not depend on  $\vec{q}_{1}$ but depend on
$\vec{k}$ and $\vec{q}_{2}.$  The transformation of these terms
into the coordinate representation by means of the integrals
(\ref{int k/k^2} )--(\ref{int I(k,q,k+q)}) and (\ref{int (kq)/kq
I(q,k,k+q)}) of Appendix A gives
\[
\tilde{g}_{p}(\vec{r}_{1},\vec{r}_{2};\vec{r}_{2}^{\;\prime})=-\frac{11}
{6}\frac{1}{\vec{r}_{12^{\prime}}^{\;2}}\ln\left(  \frac{\vec{r}_{12^{\prime}
}^{\;2}}{r_{\mu}^{2}}\right)  +\frac{11}{6}\frac{\vec{r}_{12}^{\;2}}{\vec
{r}_{22^{\prime}}^{\;2}\vec{r}_{12^{\prime}}^{\;2}}\ln\left(  \frac{\vec
{r}_{12}^{\;2}}{r_{\mu}^{2}}\right)  +\frac{11}{6}\left(  \frac{1}{\vec
{r}_{22^{\prime}}^{\,\,\,2}}-\frac{1}{\vec{r}_{12^{\prime}}^{\,\,\,2}}\right)
\ln\left(  \frac{\vec{r}_{22^{\prime}}^{\,\,\,2}}{\vec{r}_{12^{\prime}
}^{\,\,\,2}}\right)
\]
\[
+\frac{1}{4}\left(  \frac{\left(  \vec{r}_{22^{\prime}}\,\vec{r}_{12^{\prime}
}\right)  }{\vec{r}_{22^{\prime}}^{\;2}\vec{r}_{12^{\prime}}^{\;2}}-\frac
{2}{\vec{r}_{22^{\prime}}^{\;2}}\right)  \ln^{2}\left(  \frac{\vec{r}
_{12}^{\;2}}{\vec{r}_{12^{\prime}}^{\;2}}\right)  +\frac{1}{4}\left(
\frac{2\left(  \vec{r}_{22^{\prime}}\,\vec{r}_{12^{\prime}}\right)  }{\vec
{r}_{22^{\prime}}^{\;2}\vec{r}_{12^{\prime}}^{\;2}}-\frac{1}{\vec
{r}_{12^{\prime}}^{\;2}}\right)  \ln\left(  \frac{\vec{r}_{12}^{\;2}}{\vec
{r}_{22^{\prime}}^{\;2}}\right)  \ln\left(  \frac{\vec{r}_{12}^{\;2}}{\vec
{r}_{12^{\prime}}^{\;2}}\right)
\]
\begin{equation}
+\left[  \frac{\left(  \vec{r}_{22^{\prime}}\,\vec{r}_{12^{\prime}}\right)
^{\;2}}{\vec{r}_{22^{\prime}}^{\;2}\vec{r}_{12^{\prime}}^{\;2}}-1\right]
I\left(  \vec{r}_{22^{\prime}}^{\;2},\vec{r}_{12^{\prime}}^{\;2},\vec{r}
_{12}^{\;2}\right)  . \label{tilde g 0 p}
\end{equation}
The tilde is used since the expression obtained does not turn into
zero when its first two arguments coincide,  as it is required.
One can see that the only non-vanishing term for
$\vec{r}_{1}\rightarrow\vec{r}_{2}$ in this expression is the
first one. Since this term does not depend on $\vec{r}_{2},$ we
omit it to obtain
${g}_{p}(\vec{r}_{1},\vec{r}_{2};\vec{r}_{2}^{\;\prime} )$:
\begin{equation}
{g}_{p}(\vec{r}_{1},\vec{r}_{2};\vec{r}_{2}^{\;\prime})=\tilde{g}_{p}(\vec
{r}_{1},\vec{r}_{2};\vec{r}_{2}^{\;\prime})+\frac{11}{6}\frac{1}{\vec
{r}_{12^{\prime}}^{\;2}}\ln\left(  \frac{\vec{r}_{12^{\prime}}^{\;2}}{r_{\mu
}^{2}}\right)  ~. \label{g 0 p}
\end{equation}
Note that $g_{p}(\vec{r}_{1},\vec{r}_{2};\vec{r}_{2}^{\;\prime})$ has
non-integrable ultraviolet singularities at $\vec{r}_{2^{\prime}}=\vec{r}_{2}
$\ and $\vec{r}_{2^{\prime}}=\vec{r}_{1}.$\ The former singularity cancels the
corresponding singularity in $g_{p}^{0}(\vec{r}_{1},\vec{r}_{2};\rho)$\ at
$\vec{\rho}=\vec{r}_{2}$, while the latter is unessential because of the
\textquotedblleft dipole\textquotedblright\ property of \textquotedblleft
target\textquotedblright\ impact factors.

To calculate the contribution of the remaining terms in
(\ref{planar part for integration}) we need the integrals
(\ref{int (qk)/k I(q,k,q+k)}) and (\ref{int (q1q2)/q1q2 ln(q1+q2)/k}
)--(\ref{int (q1q2)/q1q2 ln q1/k}) of Appendix A and (\ref{int f4}) of
Appendix B, in addition to the integrals used before. We get
\[
\tilde{g}_{p}(\vec{r}_{1},\vec{r}_{2};\vec{r}_{1}^{\;\prime},\vec{r}
_{2}^{\;\prime})=\frac{1}{4\vec{r}_{1^{\prime}2^{\prime}}^{\;2}}\left[
\frac{(\vec{r}_{11^{\prime}}\,\vec{r}_{22^{\prime}})}{\vec{r}_{11^{\prime}
}^{\,\,2}\,\vec{r}_{22^{\prime}}^{\,\,2}}\ln\left(  \frac{\vec{r}_{21^{\prime
}}^{\,\,2}\,\vec{r}_{12^{\prime}}^{\,\,2}}{\vec{r}_{1^{\prime}2^{\prime}
}^{\,\,2}\vec{r}_{12}^{\,\,2}}\right)  +\frac{(\vec{r}_{21^{\prime}}\,\vec
{r}_{12^{\prime}})}{\vec{r}_{21^{\prime}}^{\,\,2}\,\vec{r}_{12^{\prime}
}^{\,\,2}}\ln\left(  \frac{\vec{r}_{11^{\prime}}^{\,\,2}\,\vec{r}_{22^{\prime
}}^{\,\,2}}{\vec{r}_{1^{\prime}2^{\prime}}^{\,\,2}\vec{r}_{12}^{\,\,2}
}\right)  \right]
\]
\[
-\frac{\vec{r}_{22^{\prime}}}{2\vec{r}_{22^{\prime}}^{\;2}}\left[
\frac{\,\,\vec{r}_{1^{\prime}2^{\prime}}}{\vec{r}_{1^{\prime}2^{\prime}}
^{\;2}}+\frac{\vec{r}_{11^{\prime}}}{\vec{r}_{11^{\prime}}^{\;2}}
+\frac{\left(  \vec{r}_{12^{\prime}}\,\,\vec{r}_{1^{\prime}2^{\prime}}\right)
}{\vec{r}_{1^{\prime}2^{\prime}}^{\;2}}\frac{\partial}{\partial\vec{r}_{1}
}-\frac{\left(  \vec{r}_{11^{\prime}}\,\,\vec{r}_{12^{\prime}}\right)  }
{\vec{r}_{11^{\prime}}^{\;2}}\frac{\partial}{\partial\vec{r}_{2^{\prime}}
}\right]  I\left(  \vec{r}_{11^{\prime}}^{\;2},\vec{r}_{1^{\prime}2^{\prime}
}^{\;2},\vec{r}_{12^{\prime}}^{2}\right)
\]
\[
+\frac{\vec{r}_{22^{\prime}}}{2\vec{r}_{22^{\prime}}^{\;2}\vec{r}_{11^{\prime
}}^{\;2}}\left[  \frac{\,\,\vec{r}_{12^{\prime}}}{\vec{r}_{12^{\prime}}^{\;2}
}\ln\left(  \frac{\vec{r}_{1^{\prime}2^{\prime}}^{\;4}}{\vec{r}_{11^{\prime}
}^{\;2}\vec{r}_{12^{\prime}}^{\;2}}\right)  +\frac{\,\,\vec{r}_{1^{\prime
}2^{\prime}}}{\vec{r}_{1^{\prime}2^{\prime}}^{\;2}}\ln\left(  \frac{\vec
{r}_{11^{\prime}}^{\;2}\vec{r}_{12^{\prime}}^{\;4}}{\vec{r}_{1^{\prime
}2^{\prime}}^{\;6}}\right)  \right]
\]
\begin{equation}
+\frac{\vec{r}_{22^{\prime}}}{2\vec{r}_{22^{\prime}}^{\;2}\vec{r}_{1^{\prime
}2^{\prime}}^{\;2}}\left[  \frac{\vec{r}_{21^{\prime}}\,}{\vec{r}_{1^{\prime
}2}^{\;2}}\ln\left(  \frac{\vec{r}_{12}^{\;2}}{\vec{r}_{11^{\prime}}^{\;2}
}\right)  +\frac{\vec{r}_{12^{\prime}}\,}{\vec{r}_{12^{\prime}}^{\;2}}
\ln\left(  \frac{\vec{r}_{12^{\prime}}^{\;2}}{\vec{r}_{1^{\prime}2^{\prime}
}^{\;2}}\right)  \right]  +(1\leftrightarrow2).
\end{equation}
Here and hereafter $(1\leftrightarrow2)$\ means both $1\leftrightarrow2$\ and
$1^{\prime}\leftrightarrow2^{\prime}$\ \ substitutions. The derivatives in
this equality can be calculated using the identity (\ref{identity_with_derivative}
) of Appendix B. We find
\[
\tilde{g}_{p}(\vec{r}_{1},\vec{r}_{2};\vec{r}_{1}^{\;\prime},\vec{r}
_{2}^{\;\prime})=\left[  \frac{\left(  \vec{r}_{22^{\prime}}\,\,\vec
{r}_{1^{\prime}2^{\prime}}\right)  }{\vec{r}_{11^{\prime}}^{\;2}\vec
{r}_{22^{\prime}}^{\;2}\vec{r}_{1^{\prime}2^{\prime}}^{\;2}}+\frac{\left(
\vec{r}_{22^{\prime}}\,\,\vec{r}_{12^{\prime}}\right)  \left(  \vec
{r}_{11^{\prime}}\,\,\vec{r}_{12^{\prime}}\right)  }{\vec{r}_{11^{\prime}
}^{\;2}\vec{r}_{22^{\prime}}^{\;2}\vec{r}_{1^{\prime}2^{\prime}}^{\;2}\vec
{r}_{12^{\prime}}^{\;2}}\right]  \ln\left(  \frac{\vec{r}_{12^{\prime}}^{\;2}
}{\vec{r}_{1^{\prime}2^{\prime}}^{\;2}}\right)  -\frac{\left(  \vec
{r}_{22^{\prime}}\vec{r}_{12^{\prime}}\right)  }{2\vec{r}_{22^{\prime}}
^{\;2}\vec{r}_{1^{\prime}2^{\prime}}^{\;2}\vec{r}_{12^{\prime}}^{\;2}}
\ln\left(  \frac{\vec{r}_{11^{\prime}}^{\;2}}{\vec{r}_{1^{\prime}2^{\prime}
}^{\;2}}\right)
\]
\begin{equation}
+\frac{1}{4\vec{r}_{1^{\prime}2^{\prime}}^{\;2}}\left[  \frac{(\vec
{r}_{11^{\prime}}\,\vec{r}_{22^{\prime}})}{\vec{r}_{11^{\prime}}^{\,\,2}
\,\vec{r}_{22^{\prime}}^{\,\,2}}\ln\left(  \frac{\vec{r}_{21^{\prime}}
^{\,\,2}\,\vec{r}_{12^{\prime}}^{\,\,2}}{\vec{r}_{1^{\prime}2^{\prime}
}^{\,\,2}\vec{r}_{12}^{\,\,2}}\right)  +\frac{(\vec{r}_{21^{\prime}}\,\vec
{r}_{12^{\prime}})}{\vec{r}_{21^{\prime}}^{\,\,2}\,\vec{r}_{12^{\prime}
}^{\,\,2}}\ln\left(  \frac{\vec{r}_{11^{\prime}}^{\,\,2}\,\vec{r}_{22^{\prime
}}^{\,\,2}}{\vec{r}_{1^{\prime}2^{\prime}}^{\,\,2}\vec{r}_{12}^{\,\,2}
}\right)  \right]  +\frac{\left(  \vec{r}_{22^{\prime}}\vec{r}_{21^{\prime}
}\right)  }{2\vec{r}_{22^{\prime}}^{\;2}\vec{r}_{1^{\prime}2^{\prime}}
^{\;2}\vec{r}_{21^{\prime}}^{\;2}}\ln\left(  \frac{\vec{r}_{12}^{\;2}}{\vec
{r}_{11^{\prime}}^{\;2}}\right)  +(1\leftrightarrow2).
\end{equation}
We could obtain ${g}_{p}(\vec{r}_{1},\vec{r}_{2};\vec{r}_{1}^{\;\prime}
,\vec{r}_{2}^{\;\prime})$\ subtracting from $\tilde{g}_{p}$\ the half-sum of
its values with $\vec{r}_{2}$ changed into $\vec{r}_{1}$ and vice versa. Since
${g}_{p}$ is not unique, we prefer to construct a shorter form for it. We
have
\[
{g}_{p}(\vec{r}_{1},\vec{r}_{2};\vec{r}_{1}^{\;\prime},\vec{r}_{2}^{\;\prime
})=\tilde{g}_{p}(\vec{r}_{1},\vec{r}_{2};\vec{r}_{1}^{\;\prime},\vec{r}
_{2}^{\;\prime})
\]
\[
+\left[  \frac{(\vec{r}_{22^{\prime}}\,\,\vec{r}_{21^{\prime}})}{2\vec
{r}_{1^{\prime}2^{\prime}}^{\;2}\vec{r}_{22^{\prime}}^{\;2}\vec{r}
_{21^{\prime}}^{\;2}}\ln\left(  \frac{\vec{r}_{1^{\prime}2^{\prime}}^{\;2}
}{\vec{r}_{22^{\prime}}^{\;2}}\right)  +\frac{1}{2\vec{r}_{1^{\prime}
2^{\prime}}^{\;2}\vec{r}_{12^{\prime}}^{\;2}}\ln\left(  \frac{\vec
{r}_{11^{\prime}}^{\;2}}{\vec{r}_{1^{\prime}2^{\prime}}^{\;2}}\right)
-\frac{1}{\vec{r}_{11^{\prime}}^{\;2}\vec{r}_{1^{\prime}2^{\prime}}^{\;2}}
\ln\left(  \frac{\vec{r}_{12^{\prime}}^{\;2}}{\vec{r}_{1^{\prime}2^{\prime}
}^{\;2}}\right)  +(1\leftrightarrow2)\right]
\]
\[
=\frac{1}{4\vec{r}_{1^{\prime}2^{\prime}}^{\;2}}\left[  \frac{(\vec
{r}_{11^{\prime}}\,\vec{r}_{22^{\prime}})}{\vec{r}_{11^{\prime}}^{\,\,2}
\,\vec{r}_{22^{\prime}}^{\,\,2}}+\frac{(\vec{r}_{21^{\prime}}\,\vec
{r}_{12^{\prime}})}{\vec{r}_{21^{\prime}}^{\,\,2}\,\vec{r}_{12^{\prime}
}^{\,\,2}}-\frac{2(\vec{r}_{22^{\prime}}\,\,\vec{r}_{21^{\prime}})}{\vec
{r}_{22^{\prime}}^{\;2}\vec{r}_{21^{\prime}}^{\;2}}\right]  \ln\left(
\frac{\vec{r}_{11^{\prime}}^{\,\,2}\,\vec{r}_{22^{\prime}}^{\,\,2}}{\vec
{r}_{1^{\prime}2^{\prime}}^{\,\,2}\vec{r}_{12}^{\,\,2}}\right)  +\frac
{(\vec{r}_{11^{\prime}}\,\vec{r}_{22^{\prime}})}{4\vec{r}_{11^{\prime}
}^{\,\,2}\,\vec{r}_{22^{\prime}}^{\,\,2}\vec{r}_{1^{\prime}2^{\prime}}^{\;2}
}\ln\left(  \frac{\vec{r}_{21^{\prime}}^{\,\,2}\,\vec{r}_{12^{\prime}}
^{\,\,2}}{\vec{r}_{11^{\prime}}^{\,\,2}\,\vec{r}_{22^{\prime}}^{\,\,2}
}\right)
\]
\[
-\frac{(\vec{r}_{12}\,\,\vec{r}_{22^{\prime}})}{2\vec{r}_{1^{\prime}2^{\prime
}}^{\;2}\vec{r}_{22^{\prime}}^{\;2}\vec{r}_{12^{\prime}}^{\;2}}\ln\left(
\frac{\vec{r}_{11^{\prime}}^{\;2}}{\vec{r}_{1^{\prime}2^{\prime}}^{\;2}
}\right)  +\left[  \frac{\left(  \vec{r}_{22^{\prime}}\,\,\vec{r}_{12}\right)
}{\vec{r}_{11^{\prime}}^{\;2}\vec{r}_{22^{\prime}}^{\;2}\vec{r}_{1^{\prime
}2^{\prime}}^{\;2}}-\frac{\left(  \vec{r}_{22^{\prime}}\,\,\vec{r}
_{11^{\prime}}\right)  }{\vec{r}_{11^{\prime}}^{\;2}\vec{r}_{22^{\prime}
}^{\;2}\vec{r}_{1^{\prime}2^{\prime}}^{\;2}}+\frac{\left(  \vec{r}
_{22^{\prime}}\,\,\vec{r}_{12^{\prime}}\right)  \left(  \vec{r}_{11^{\prime}
}\,\,\vec{r}_{12^{\prime}}\right)  }{\vec{r}_{11^{\prime}}^{\;2}\vec
{r}_{22^{\prime}}^{\;2}\vec{r}_{1^{\prime}2^{\prime}}^{\;2}\vec{r}
_{12^{\prime}}^{\;2}}\right]  \ln\left(  \frac{\vec{r}_{12^{\prime}}^{\;2}
}{\vec{r}_{1^{\prime}2^{\prime}}^{\;2}}\right)
\]
\begin{equation}
+ \, (1\leftrightarrow2).
\end{equation}
The dipole property of this term is explicit.

\section{Transformation of the ``symmetric'' part}

\label{sec:symmetric}

In the coordinate representation the piece $\hat{\mathcal{K}}_{s1}$, see 
Eq.~(\ref{Ks1}), is given by the integral
\begin{equation}
\langle\vec{r}_{1}\vec{r}_{2}|\hat{\mathcal{K}}_{s1}|\vec{r}_{1}^{\;\prime
}\vec{r}_{2}^{\;\prime}\rangle=\frac{\alpha_{s}^{2}(\mu)N_{c}^{2}}{4\pi^{4}
}\int_{0}^{1}{dx}\int\frac{d\vec{q}_{1}}{2\pi}\frac{d\vec{q}_{2}}{2\pi}
\frac{d\vec{k}_{1}}{2\pi}\frac{d\vec{k}_{2}}{2\pi}\frac{e^{i[\vec{q}_{1}
\vec{r}_{11^{\prime}}+\vec{q}_{2}\vec{r}_{22^{\prime}}+\vec{k}\vec
{r}_{1^{\prime}2^{\prime}}]}}{\vec{q}_{1}^{\;2}\vec{q}_{2}^{\;2}}\left(
\frac{F_{s}(k_{1},k_{2})}{x(1-x)}\right)  _{+}\;. \label{integral for K s1}
\end{equation}
Here $F_{s}(k_{1},k_{2})$ is given by Eqs.~(3.44), (3.45) and (4.1) of
Ref.~\cite{FF05} and $\vec{k}=\vec{k}_{1}+\vec{k}_{2}$. We restrict ourselves
to the dipole form of the kernel. Hence, we omit those terms in $F_{s}
(k_{1},k_{2}),$ which lead to $\delta(\vec{r}_{1^{\prime}2^{\prime}})$ in the
coordinate representation. Decomposing the remaining terms and taking into
account the symmetry of the integration measure in (\ref{integral for K s1})
with regard to the substitution $k_{1}\leftrightarrow k_{2}$, we can make
the replacement
\begin{equation}
\frac{F_{s}(k_{1},k_{2})}{\vec{q}_{1}^{\;2}\vec{q}_{2}^{\;2}}\rightarrow
-2\frac{k_{1}^{i}k_{2}^{j}}{\vec{k}_{1}^{\;2}\vec{k}_{2}^{\;2}}\frac{x_{2}
}{\sigma_{22}}a_{2}^{ij}-2\frac{x_{1}}{\sigma_{11}}a_{1}^{ij}\frac{k_{1}
^{i}k_{2}^{j}}{\vec{k}_{1}^{\;2}\vec{k}_{2}^{\;2}}+2\frac{x_{1}}{\sigma_{11}
}a_{1}^{ij}\frac{x_{2}}{\sigma_{22}}a_{2}^{ij}\;, \label{decomposition of F s}
\end{equation}
where
\[
x_{1}=x\;,\;\;\;x_{2}=1-x\;,\;\;\;\sigma_{11}=(\vec{k}_{1}-x_{1}\vec{q}
_{1})^{2}+x_{1}x_{2}\vec{q}_{1}^{\;2}\;,\;\;\;\sigma_{22}=(\vec{k}_{2}
+x_{2}\vec{q}_{2})^{2}+x_{1}x_{2}\vec{q}_{2}^{\;2}\;,
\]
and
\[
a_{1}^{ij}=\frac{\delta^{ij}}{2}x_{2}\left(  1-2\frac{(\vec{q}_{1}\vec{k}
_{1})}{\vec{q}_{1}^{\;2}}\right)  +\frac{x_{2}}{x_{1}}\frac{k_{1}^{i}q_{1}
^{j}}{\vec{q}_{1}^{\;2}}-\frac{q_{1}^{i}(q_{1}-k_{1})^{j}}{\vec{q}_{1}^{\;2}
}+\frac{k_{1}^{i}(q_{1}-k_{1})^{j}}{\vec{k}_{1}^{\;2}}\;,
\]
\begin{equation}
a_{2}^{ij}=\frac{\delta^{ij}}{2}x_{1}\left(  1+2\frac{(\vec{q}_{2}\vec{k}
_{2})}{\vec{q}_{2}^{\;2}}\right)  -\frac{x_{1}}{x_{2}}\frac{q_{2}^{i}k_{2}
^{j}}{\vec{q}_{2}^{\;2}}-\frac{(q_{2}+k_{2})^{i}q_{2}^{j}}{\vec{q}_{2}^{\;2}
}-\frac{(q_{2}+k_{2})^{i}k_{2}^{j}}{\vec{k}_{2}^{\;2}}~. \label{a ij}
\end{equation}
Since the first term in (\ref{decomposition of F s}) does not depend on
$\vec{q}_{1},$\ the second is independent of $\vec{q}_{2}$\ and the third one
depends on all momenta, they contribute to $g(\vec{r}_{1},\vec{r}_{2};\vec
{r}_{2}^{\;\prime}),$ $g(\vec{r}_{2},\vec{r}_{1};\vec{r}_{1}^{\;\prime}
)$ and $g(\vec{r}_{1},\vec{r}_{2};\vec{r}_{1}^{\,\prime};\vec{r}
_{2}^{\;\prime}) $\ in (\ref{kernel through g}), respectively. We denote these
contributions with the $_{s1}$ subscript. For the first term we have
\[
-2\frac{k_{1}^{i}k_{2}^{j}}{\vec{k}_{1}^{\;2}\vec{k}_{2}^{\;2}}\left(
\frac{x_{2}a_{2}^{ij}}{\sigma_{22}x(1-x)}\right)  _{+}=-\frac{k_{1}^{i}}
{\vec{k}_{1}^{\;2}}\left[  \frac{(k_{2}+q_{2})^{i}}{x_{1}}\left(  \frac
{1}{\vec{q}_{2}^{\;2}}-\frac{1}{\vec{k}_{2}^{\;2}}\right)  \left(  \frac
{1}{\sigma_{22}}-\frac{1}{(\vec{k}_{2}+\vec{q}_{2})^{2}}\right)  \right.
\]
\begin{equation}
\left.  -\frac{q_{2}^{i}}{x_{2}\vec{q}_{2}^{\;2}}\left(  \frac{1}{\sigma_{22}
}-\frac{1}{\vec{k}_{2}^{\;2}}\right)  \right]  ~. \label{a 02}
\end{equation}
Note, that infrared singularities vanish here, as well as in other terms in
the R.H.S. of (\ref{decomposition of F s}), in a rather tricky way, namely by
means of the $_{+}$ prescription. Indeed, each term in
(\ref{decomposition of F s}) contains non-integrable infrared singularities.
But these terms do not depend on $x$ so that the singularities vanish after
the subtraction.

For the term in question it seems more convenient to perform first the
integration over $x$. We obtain
\begin{equation}
-2\int_{0}^{1}dx\frac{k_{1}^{i}k_{2}^{j}}{\vec{k}_{1}^{\;2}\vec{k}_{2}^{\;2}
}\left(  \frac{x_{2}a_{2}^{ij}}{\sigma_{22}x(1-x)}\right)  _{+}=-\frac
{k_{1}^{i}}{\vec{k}_{1}^{\;2}}\left[  \frac{(k_{2}+q_{2})^{i}}{(\vec{k}
_{2}+\vec{q}_{2})^{2}}\left(  \frac{1}{\vec{q}_{2}^{\;2}}-\frac{1}{\vec{k}
_{2}^{\;2}}\right)  +\frac{q_{2}^{i}}{\vec{q}_{2}^{\;2}\vec{k}_{2}^{\;2}
}\right]  \ln\left(  \frac{(\vec{k}_{2}+\vec{q}_{2})^{2}}{\vec{k}_{2}^{\;2}
}\right)  ~. \label{a 02 integrated over x}
\end{equation}
Then we use the integrals (\ref{int p/qp ln p/(q+p)}) and
(\ref{int (exp-exp)q/qk ln q/k}) of Appendix B to find the contribution of
this term to $g_{s1}(\vec{r}_{1},\vec{r}_{2};\vec{r}_{2}^{\;\prime}).$
\begin{equation}
g_{s1}(\vec{r}_{1},\vec{r}_{2};\vec{r}_{2}^{\;\prime})=\left[  1-\frac
{(\vec{r}_{22^{\prime}}\,\vec{r}_{12^{\prime}})^{\;2}}{\vec{r}_{22^{\prime}
}^{\,2}\vec{r}_{12^{\prime}}^{\,2}}\right]  I(\vec{r}_{12}^{\;2},\,\vec
{r}_{12^{\prime}}^{\;2},\vec{r}_{22^{\prime}}^{\;2})+\left(  \frac{(\vec
{r}_{22^{\prime}}\,\vec{r}_{12^{\prime}})}{2\vec{r}_{22^{\prime}}^{\,2}\vec
{r}_{12^{\prime}}^{\,2}}-\frac{1}{2\vec{r}_{12^{\prime}}^{\,2}}\right)
\ln\frac{\vec{r}_{12}^{\;2}}{\vec{r}_{22^{\prime}}^{\;2}}\ln\frac{\vec{r}
_{12}^{\;2}}{\vec{r}_{12^{\prime}}^{\;2}}. \label{F s first part integrated}
\end{equation}
One can see that the second term in (\ref{decomposition of F s}) equals the
first one after the substitution $\vec{q}_{1}\rightarrow-\vec{q}_{2},\,$
$\vec{k}_{1}\rightarrow\vec{k}_{2},\,$ $x_{1}\rightarrow x_{2}.$\ Hence, we
can construct its integrated form replacing $\vec{r}_{22^{\prime}}
\rightarrow-\vec{r}_{11^{\prime}},$ $\vec{r}_{1^{\prime}2}\rightarrow\vec
{r}_{12^{\prime}},$ $\vec{r}_{12^{\prime}}\rightarrow\vec{r}_{1^{\prime}2}
$\ in\ (\ref{F s first part integrated}). The third term in
(\ref{decomposition of F s}) is easier to integrate with respect to momenta
before the convolution. We introduce
\[
\vec{l}_{1}=\vec{k}_{1}-x\vec{q}_{1},\quad\vec{l}_{2}=\vec{k}_{2}+(1-x)\vec
{q}_{2},\quad\vec{p}_{1}=\vec{q}_{1}-\vec{k}_{1},\quad\vec{p}_{2}=\vec{q}
_{2}+\vec{k}_{2}.
\]
In this notation
\[
\left(  \frac{2x_{1}a_{1}^{ij}x_{2}a_{2}^{ij}}{x(1-x)\sigma_{11}\sigma_{22}
}\right)  _{+}=2\frac{1}{\sigma_{11}}\left[  \frac{\delta^{ij}}{2}x_{2}\left(
(1-2x_{1})-2\frac{(\vec{q}_{1}\,\vec{l}_{1})}{\vec{q}_{1}^{\;2}}\right)
+\frac{x_{2}}{x_{1}}\frac{l_{1}^{i}q_{1}^{j}}{\vec{q}_{1}^{\;2}}+\frac
{l_{1}^{j}q_{1}^{i}}{\vec{q}_{1}^{\;2}}+\frac{k_{1}^{i}p_{1}^{j}}{\vec{k}
_{1}^{\;2}}\right]
\]
\[
\times\frac{1}{\sigma_{22}}\left[  \frac{\delta^{ij}}{2}x_{1}\left(
(1-2x_{2})+2\frac{(\vec{q}_{2}\,\vec{l}_{2})}{\vec{q}_{2}^{\;2}}\right)
-\frac{x_{1}}{x_{2}}\frac{l_{2}^{j}q_{2}^{i}}{\vec{q}_{2}^{\;2}}-\frac
{l_{2}^{i}q_{2}^{j}}{\vec{q}_{2}^{\;2}}-\frac{k_{2}^{j}p_{2}^{i}}{\vec{k}
_{2}^{\;2}}\right]
\]
\begin{equation}
+\frac{2k_{1}^{i}q_{1}^{j}}{x_{1}\vec{k}_{1}^{\;2}\vec{q}_{1}^{\;2}}\left(
\frac{q_{2}^{j}}{\vec{q}_{2}^{\;2}}+\frac{k_{2}^{j}}{\vec{k}_{2}^{\;2}
}\right)  \frac{(k_{2}+q_{2})^{i}}{(k_{2}+q_{2})^{2}}+\frac{2k_{2}^{j}
q_{2}^{i}}{x_{2}\vec{k}_{2}^{\;2}\vec{q}_{2}^{\;2}}\left(  \frac{q_{1}^{i}
}{\vec{q}_{1}^{\;2}}-\frac{k_{1}^{i}}{\vec{k}_{1}^{\;2}}\right)  \frac
{(k_{1}-q_{1})^{j}}{(k_{1}-q_{1})^{2}}\;.
\end{equation}
Then we use the integrals (\ref{int 1/sigma})--(\ref{int (qk)/(k
sigma)}) of Appendix B to find
\[
\int\frac{d^{2}q_{1}}{2\pi}\frac{d^{2}q_{2}}{2\pi}\frac{d^{2}k_{1}}{2\pi}
\frac{d^{2}k_{2}}{2\pi}\left(  \frac{2x_{1}a_{1}^{ij}x_{2}a_{2}^{ij}
}{x(1-x)\sigma_{11}\sigma_{22}}\right)  _{+}e^{i[\vec{q}_{1}\,\vec
{r}_{11^{\prime}}+\vec{q}_{2}\,\vec{r}_{22^{\prime}}+\vec{k}\,\vec
{r}_{1^{\prime}2^{\prime}}^{\;}]}
\]
\[
=\frac{2}{d_{1}d_{2}}\left[  \frac{\delta^{ij}}{2}x_{2}\left(  1+2\frac
{(\vec{r}_{11^{\prime}}\,\vec{r}_{1^{\prime}2^{\prime}})}{\vec{r}_{1^{\prime
}2^{\prime}}^{\;2}}\right)  -\frac{r_{12^{\prime}}^{i}r_{1^{\prime}2^{\prime}
}^{j}}{\vec{r}_{1^{\prime}2^{\prime}}^{\;2}}-\frac{x_{2}}{x_{1}}
\frac{r_{11^{\prime}}^{j}r_{1^{\prime}2^{\prime}}^{i}}{\vec{r}_{1^{\prime
}2^{\prime}}^{\;2}}-\frac{r_{12^{\prime}}^{i}r_{11^{\prime}}^{j}}{\vec
{r}_{11^{\prime}}^{\;2}}\right]
\]
\[
\times\left[  \frac{\delta^{ij}}{2}x_{1}\left(  1-2\frac{(\vec{r}_{22^{\prime
}}\,\vec{r}_{1^{\prime}2^{\prime}})}{\vec{r}_{1^{\prime}2^{\prime}}^{\;2}
}\right)  +\frac{r_{21^{\prime}}^{j}r_{1^{\prime}2^{\prime}}^{i}}{\vec
{r}_{1^{\prime}2^{\prime}}^{\;2}}+\frac{x_{1}}{x_{2}}\frac{r_{22^{\prime}}
^{i}r_{1^{\prime}2^{\prime}}^{j}}{\vec{r}_{1^{\prime}2^{\prime}}^{\;2}}
-\frac{r_{21^{\prime}}^{j}r_{22^{\prime}}^{i}}{\vec{r}_{22^{\prime}}^{\;2}
}\right]
\]
\begin{equation}
+\frac{2}{x_{1}}\frac{(\vec{r}_{11^{\prime}}\,\,\vec{r}_{21^{\prime}})}
{\vec{r}_{11^{\prime}}^{\;2}\vec{r}_{21^{\prime}}^{\;2}}\frac{(\vec
{r}_{22^{\prime}}\,\vec{r}_{21^{\prime}})}{\vec{r}_{22^{\prime}}^{\;2}\vec
{r}_{1^{\prime}2^{\prime}}^{\;2}}+\frac{2}{x_{2}}\frac{(\vec{r}_{22^{\prime}
}\,\vec{r}_{12^{\prime}})}{\vec{r}_{22^{\prime}}^{\;2}\vec{r}_{12^{\prime}
}^{\;2}}\frac{(\vec{r}_{11^{\prime}}\,\vec{r}_{12^{\prime}})}{\vec
{r}_{11^{\prime}}^{\;2}\vec{r}_{1^{\prime}2^{\prime}}^{\;2}}\;,
\end{equation}
where $d_{1}=x_{1}\vec{r}_{12^{\prime}}^{\;2}+x_{2}\vec{r}_{11^{\prime}}
^{\;2}$ and $d_{2}=x_{1}\vec{r}_{22^{\prime}}^{\;2}+x_{2}\vec{r}_{21^{\prime}
}^{\;2}$. The subsequent integration over $x$ is straightforward. We obtain
\[
\tilde{g}_{s1}(\vec{r}_{1},\vec{r}_{2};\vec{r}_{1}^{\,\prime},\vec{r}
_{2}^{\;\prime})={g}_{s1}(\vec{r}_{1},\vec{r}_{2};\vec{r}_{1}^{\,\prime}
,\vec{r}_{2}^{\;\prime})-\frac{1}{\vec{r}_{1^{\prime}2^{\prime}}^{\,\,4}
}\left[  \frac{\vec{r}_{11^{\prime}}^{\;2}}{(\vec{r}_{12^{\prime}}^{\;2}
-\vec{r}_{11^{\prime}}^{\;2})}\ln\frac{\vec{r}_{12^{\prime}}^{\;2}}{\vec
{r}_{11^{\prime}}^{\;2}}+\frac{\vec{r}_{22^{\prime}}^{\;2}}{(\vec
{r}_{21^{\prime}}^{\;2}-\vec{r}_{22^{\prime}}^{\;2})}\ln\frac{\vec
{r}_{21^{\prime}}^{\;2}}{\vec{r}_{22^{\prime}}^{\;2}}-2\right]
\]
\qquad
\begin{equation}
+\frac{\vec{r}_{11^{\prime}}^{\;2}+(\vec{r}_{11^{\prime}}\,\vec{r}
_{12^{\prime}})}{\,(\vec{r}_{12^{\prime}}^{\;2}-\vec{r}_{11^{\prime}}
^{\;2})\,\vec{r}_{1^{\prime}2^{\prime}}^{\;2}\vec{r}_{11^{\prime}}^{\;2}\,}
\ln\frac{\vec{r}_{12^{\prime}}^{\;2}\,}{\vec{r}_{11^{\prime}}^{\;2}}
+\frac{\vec{r}_{22^{\prime}}^{\;2}+(\vec{r}_{22^{\prime}}\,\vec{r}
_{21^{\prime}})}{\,\,(\vec{r}_{21^{\prime}}^{\;2}-\vec{r}_{22^{\prime}}
^{\;2})\,\vec{r}_{1^{\prime}2^{\prime}}^{\;2}\vec{r}_{22^{\prime}}^{\;2}}
\ln\frac{\vec{r}_{21^{\prime}}^{\;2}}{\vec{r}_{22^{\prime}}^{\;2}}.
\label{F s third part integrated}
\end{equation}
Here
\[
{g}_{s1}(\vec{r}_{1},\vec{r}_{2};\vec{r}_{1}^{\,\prime},\vec{r}_{2}^{\;\prime
})=\frac{1}{\vec{r}_{1^{\prime}2^{\prime}}^{\,\,4}}\left(  \frac{\vec
{r}_{11^{\prime}}^{\;2}\,\vec{r}_{22^{\prime}}^{\;2}}{d}\ln\left(  \frac
{\vec{r}_{12^{\prime}}^{\;2}\,\vec{r}_{21^{\prime}}^{\;2}}{\vec{r}
_{11^{\prime}}^{\;2}\vec{r}_{22^{\prime}}^{\;2}}\right)  -1\right)  +\frac
{1}{d\,\vec{r}_{1^{\prime}2^{\prime}}^{\;2}}\left[  \frac{(\vec{r}_{1^{\prime
}2^{\prime}}\,\vec{r}_{12^{\prime}})\vec{r}_{12}^{\;2}}{\vec{r}_{11^{\prime}
}^{\;2}}-\frac{(\vec{r}_{1^{\prime}2^{\prime}}\,\vec{r}_{21^{\prime}})\vec
{r}_{12}^{\;2}}{\vec{r}_{22^{\prime}}^{\;2}}\right.
\]
\qquad
\[
\left.  +\frac{2(\vec{r}_{22^{\prime}}\,\vec{r}_{21^{\prime}})(\vec{r}
_{12}\,\vec{r}_{21^{\prime}})}{\vec{r}_{21^{\prime}}^{\;2}}-\frac{2(\vec
{r}_{11^{\prime}}\,\,\vec{r}_{12^{\prime}})(\vec{r}_{12}\,\vec{r}_{12^{\prime
}})}{\vec{r}_{12^{\prime}}^{\;2}}+\frac{2(\vec{r}_{22^{\prime}}\,\vec
{r}_{12^{\prime}})(\vec{r}_{11^{\prime}}\,\vec{r}_{21^{\prime}})}{\vec
{r}_{11^{\prime}}^{\;2}\,\vec{r}_{22^{\prime}}^{\;2}}\vec{r}_{1^{\prime
}2^{\prime}}^{\;2}-2\vec{r}_{1^{\prime}2^{\prime}}^{\;2}\right]  \ln\left(
\frac{\vec{r}_{12^{\prime}}^{\;2}\,\vec{r}_{21^{\prime}}^{\;2}}{\vec
{r}_{11^{\prime}}^{\;2}\,\vec{r}_{22^{\prime}}^{\;2}}\right)
\]
\[
+\frac{1}{\vec{r}_{22^{\prime}}^{\;2}}\left(  -\,\frac{(\vec{r}_{12}\,\vec
{r}_{12^{\prime}})}{\vec{r}_{1^{\prime}2^{\prime}}^{\;2}\,\vec{r}_{12^{\prime
}}^{\;2}}+\frac{(\vec{r}_{22^{\prime}}\,\vec{r}_{12})}{\vec{r}_{1^{\prime
}2^{\prime}}^{\;2}\vec{r}_{11^{\prime}}^{\;2}}-\frac{(\vec{r}_{22^{\prime}
}\,\vec{r}_{12^{\prime}})}{\vec{r}_{11^{\prime}}^{\;2}\,\vec{r}_{12^{\prime}
}^{\;2}}\right)  \ln\frac{\vec{r}_{21^{\prime}}^{\;2}}{\vec{r}_{22^{\prime}
}^{\;2}}
\]
\begin{equation}
+\frac{1}{\vec{r}_{11^{\prime}}^{\;2}\,}\left(  \frac{(\vec{r}_{12}\,\vec
{r}_{21^{\prime}})}{\vec{r}_{1^{\prime}2^{\prime}}^{\;2}\vec{r}_{21^{\prime}
}^{\;2}}-\frac{(\vec{r}_{11^{\prime}}\,\vec{r}_{12})}{\vec{r}_{1^{\prime
}2^{\prime}}^{\;2}\vec{r}_{22^{\prime}}^{\;2}}-\frac{(\vec{r}_{11^{\prime}
}\,\vec{r}_{21^{\prime}})}{\vec{r}_{22^{\prime}}^{\;2}\vec{r}_{21^{\prime}
}^{\;2}}\right)  \ln\frac{\vec{r}_{12^{\prime}}^{\;2}\,}{\vec{r}_{11^{\prime}
}^{\;2}}, \label{F s third part final}
\end{equation}
where
$d=\vec{r}_{12^{\prime}}^{\;2}\vec{r}_{21^{\prime}}^{\;2}-\vec
{r}_{11^{\prime}}^{\;2}\vec{r}_{22^{\prime}}^{\;2}.$ One can see
that ${g}_{s1}$\ vanishes when $\vec{r}_{1}$  equals $\vec{r}_{2}$
while the remaining terms in (\ref{F s third part integrated}) are
independent of either $\vec{r}_{1}$ or $\vec{r}_{2}$, therefore we
dropped them.

Now we turn to $\hat{\mathcal{K}}_{s2}$. We can rewrite this expression in a
form convenient for the integration
\[
\langle\vec{r}_{1},\vec{r}_{2}|\hat{\mathcal{K}}_{s2}|\vec{r}_{1}^{\;\prime
},\vec{r}_{2}^{\;\prime}\rangle=-\frac{\alpha_{s}^{2}(\mu)N_{c}^{2}}{4\pi^{4}
}\int\frac{d\vec{q}_{1}}{2\pi}\frac{d\vec{q}_{2}}{2\pi}\frac{d\vec{k}_{1}
}{2\pi}\frac{d\vec{k}_{2}}{2\pi}\ln\left(  \frac{\vec{k}_{2}^{\;2}}{\vec
{k}_{1}^{\;2}}\right)  e^{i[\vec{q}_{1}\,\vec{r}_{11^{\prime}}+\vec{q}
_{2}\,\vec{r}_{22^{\prime}}+\vec{k}\,\,\vec{r}_{1^{\prime}2^{\prime}}]}\;
\]
\[
\left[  \frac{1}{\vec{k}_{1}^{\;2}\vec{k}_{2}^{\;2}}+\frac{1}{\vec{k}
_{1}^{\;2}}\left\{  \frac{\vec{k}_{2}}{\vec{k}_{2}^{\;2}}\left(  \frac{\vec
{q}_{2}+\vec{k}_{1}}{(\vec{q}_{2}+\vec{k}_{1})^{2}}-\frac{\vec{q}_{2}}{\vec
{q}_{2}^{\;2}}-\frac{\vec{q}_{1}-\vec{k}_{1}}{(q_{1}-\vec{k}_{1})^{2}}
+\frac{\vec{q}_{1}}{\vec{q}_{1}^{\;2}}\right)  -\frac{(\vec{q}_{1}-\vec{k}
_{1})(\vec{q}_{2}+\vec{k}_{1})}{(\vec{q}_{1}-\vec{k}_{1})^{2}(\vec{q}_{2}
+\vec{k}_{1})^{2}}+\frac{\left(  \vec{q}_{1}\,\vec{q}_{2}\right)  }{\vec
{q}_{1}^{\;2}\vec{q}_{2}^{\;2}}\right\}  \right.
\]
\[
+\left(  \frac{\vec{k}_{1}}{\vec{k}_{1}^{\;2}}\left(  \frac{\vec{q}_{2}}
{\vec{q}_{2}^{\;2}}-\frac{\vec{q}_{1}}{\vec{q}_{1}^{\;2}}\right)  \right)
\left(  \frac{\vec{k}_{2}}{\vec{k}_{2}^{\;2}}\left(  \frac{\vec{q}_{2}+\vec
{k}_{1}}{(\vec{q}_{2}+\vec{k}_{1})^{2}}-\frac{\vec{q}_{1}-\vec{k}_{1}}
{(\vec{q}_{1}-\vec{k}_{1})^{2}}\right)  -\frac{(\vec{q}_{1}-\vec{k}_{1}
)(\vec{q}_{2}+\vec{k}_{1})}{(\vec{q}_{1}-\vec{k}_{1})^{2}(\vec{q}_{2}+\vec
{k}_{1})^{2}}\right)
\]
\begin{equation}
-\left.  \frac{\left(  \vec{q}_{1}\,\vec{q}_{2}\right)  }{\vec{q}_{1}
^{\;2}\vec{q}_{2}^{\;2}}\left(  \frac{\vec{k}_{2}}{\vec{k}_{2}^{\;2}}\left(
\frac{\vec{q}_{2}+\vec{k}_{1}}{(\vec{q}_{2}+\vec{k}_{1})^{2}}-\frac{\vec
{q}_{1}-\vec{k}_{1}}{(\vec{q}_{1}-\vec{k}_{1})^{2}}\right)  -\frac{(\vec
{q}_{1}-\vec{k}_{1})(\vec{q}_{2}+\vec{k}_{1})}{(\vec{q}_{1}-\vec{k}_{1}
)^{2}(\vec{q}_{2}+\vec{k}_{1})^{2}}\right)  \right]  .
\end{equation}
The first term in the square brackets vanishes in the integration.
Again, we omit terms proportional to
$\delta(\vec{r}_{1^{\prime}2^{\prime}})$ in the coordinate
representation. The remaining  terms which do not depend on
$\vec{q}_{1} $ contribute to
$g_{s2}(\vec{r}_{1},\vec{r}_{2};\vec{r}_{2}^{\;\prime}), $ the
ones independent of $\vec{q}_{2}$ contribute to
$g_{s2}(\vec{r}_{2},\vec {r}_{1};\vec{r}_{1}^{\;\prime})$ and the
 others  to $g_{s2}(\vec
{r}_{1},\vec{r}_{2};\vec{r}_{1}^{\;\prime},\vec{r}_{2}^{\;\prime}).$
During the calculation we used the integrals (\ref{int
(exp-exp)q/qk ln q/k}) and (\ref{int (exp-exp)1/k})--(\ref{int
f4}) of Appendix B. Finally, after the integration we obtain
\begin{equation}
g_{s2}(\vec{r}_{1},\vec{r}_{2};\vec{r}_{2}^{\;\prime})=\frac{1}{4\vec
{r}_{12^{\prime}}^{\;2}}\ln\left(  \frac{\vec{r}_{12}^{\;2}}{\vec
{r}_{22^{\prime}}^{\,2}}\right)  \ln\left(  \frac{\vec{r}_{12}^{\;2}}{\vec
{r}_{12^{\prime}}^{\;2}}\right)  -\frac{(\vec{r}_{12^{\prime}}\,\,\vec
{r}_{22^{\prime}})}{4\vec{r}_{12^{\prime}}^{\;2}\vec{r}_{22^{\prime}}^{\;2}
}\ln^{2}\left(  \frac{\vec{r}_{12}^{\;2}}{\vec{r}_{12^{\prime}}^{\;2}}\right)
, \label{g s2(r1 r2 r2')}
\end{equation}

\begin{equation}
\tilde{g}_{s2}(\vec{r}_{1},\vec{r}_{2};\vec{r}_{1}^{\;\prime},\vec{r}
_{2}^{\;\prime})=\tilde{g}_{p}(\vec{r}_{1},\vec{r}_{2};\vec{r}_{1}^{\;\prime
},\vec{r}_{2}^{\;\prime})-\left[  \left(  \frac{\left(  \vec{r}_{22^{\prime}
}\,\,\vec{r}_{12}\right)  }{\vec{r}_{11^{\prime}}^{\;2}\vec{r}_{22^{\prime}
}^{\;2}\vec{r}_{1^{\prime}2^{\prime}}^{\;2}}+\frac{1}{\vec{r}_{11^{\prime}
}^{\;2}\vec{r}_{1^{\prime}2^{\prime}}^{\;2}}\right)  \ln\left(  \frac{\vec
{r}_{12^{\prime}}^{\;2}}{\vec{r}_{1^{\prime}2^{\prime}}^{\;2}}\right)
+(1\leftrightarrow2)\right]  .
\end{equation}
Hence,
\begin{equation}
g_{s2}(\vec{r}_{1},\vec{r}_{2};\vec{r}_{1}^{\;\prime},\vec{r}_{2}^{\;\prime
})=g_{p}(\vec{r}_{1},\vec{r}_{2};\vec{r}_{1}^{\;\prime},\vec{r}_{2}^{\;\prime
})-\left[  \frac{\left(  \vec{r}_{22^{\prime}}\,\,\vec{r}_{12}\right)  }
{\vec{r}_{11^{\prime}}^{\;2}\vec{r}_{22^{\prime}}^{\;2}\vec{r}_{1^{\prime
}2^{\prime}}^{\;2}}\ln\left(  \frac{\vec{r}_{12^{\prime}}^{\;2}}{\vec
{r}_{1^{\prime}2^{\prime}}^{\;2}}\right)  +(1\leftrightarrow2)\right]  .
\label{g s2(r1 r2 r1' r2')}
\end{equation}
Here we used the \textquotedblleft dipole\textquotedblright\ form of $g_{p}
$\ and dropped the terms independent of $\vec{r}_{1}$\ or of $\vec{r}_{2}.$

\section{Final result}

\label{sec:result}

In this Section we gather the contributions from the \textquotedblleft
planar\textquotedblright\ and \textquotedblleft symmetric\textquotedblright
\ parts of the kernel to the functions $g$ defined in (\ref{kernel through g}).
We have
\begin{equation}
g^{0}(\vec{r}_{1},\vec{r}_{2};\rho)=\frac{3}{2}\frac{\,\vec{r}_{12}^{\;2}
}{\vec{r}_{1\rho}^{\;2}\vec{r}_{2\rho}^{\;2}}\ln\left(  \frac
{\vec{r}_{1\rho}^{\;2}}{\vec{r}_{12}^{\;2}}\right)  \ln\left(
\frac{\vec {r}_{2\rho}^{\;2}}{\vec{r}_{12}^{\;2}}\right)
-\frac{11}{12}\left[
\frac{\vec{r}_{12}^{\;2}}{\vec{r}_{1\rho}^{\;2}\vec{r}_{2\rho}^{\;2}}
\ln\left(
\frac{\vec{r}_{1\rho}^{\;2}\vec{r}_{2\rho}^{\;2}}{r_{\mu}^{4}
}\right)  +\left(
\frac{1}{\vec{r}_{2\rho}^{\,\,\,2}}-\frac{1}{\vec{r}
_{1\rho}^{\,\,\,2}}\right)  \ln\left(
\frac{\vec{r}_{2\rho}^{\,\,\,2}}
{\vec{r}_{1\rho}^{\,\,\,2}}\right)  \right]  ~,
\end{equation}
\[
g(\vec{r}_{1},\vec{r}_{2};\vec{r}_{2}^{\;\prime})\ =\frac{11}{6}\frac{\vec
{r}_{12}^{\;2}}{\vec{r}_{22^{\prime}}^{\;2}\vec{r}_{12^{\prime}}^{\;2}}
\ln\left(  \frac{\vec{r}_{12}^{\;2}}{r_{\mu}^{2}}\right)  +\frac{11}{6}\left(
\frac{1}{\vec{r}_{22^{\prime}}^{\,\,\,2}}-\frac{1}{\vec{r}_{12^{\prime}
}^{\,\,\,2}}\right)  \ln\left(  \frac{\vec{r}_{22^{\prime}}^{\,\,\,2}}{\vec
{r}_{12^{\prime}}^{\,\,\,2}}\right)
\]
\begin{equation}
+\frac{1}{2\vec{r}_{22^{\prime}}^{\;2}}\ln\left(  \frac{\vec{r}_{12^{\prime}
}^{\;2}}{\vec{r}_{22^{\prime}}^{\;2}}\right)  \ln\left(  \frac{\vec{r}
_{12}^{\;2}}{\vec{r}_{12^{\prime}}^{\;2}}\right)  -\frac{\vec{r}_{12}^{\;2}
}{2\,\vec{r}_{22^{\prime}}^{\;2}\vec{r}_{12^{\prime}}^{\;2}}\ln\left(
\frac{\vec{r}_{12}^{\;2}}{\vec{r}_{22^{\prime}}^{\;2}}\right)  \ln\left(
\frac{\vec{r}_{12}^{\;2}}{\vec{r}_{12^{\prime}}^{\;2}}\right)  ,
\end{equation}
where
\begin{equation}
\ln r_{\mu}^{2}=2\psi(1)-\ln\frac{\mu^{2}}{4}-\frac{3}{11}\left(  \frac{67}
{9}-2\zeta(2)\right)  .
\end{equation}
One can see that both $g^{0}(\vec{r}_{1},\vec{r}_{2};\vec{\rho})$ and
$g(\vec{r}_{1},\vec{r}_{2};\vec{\rho})$ vanish at $\vec{r}_{1}=\vec{r}_{2}$.
Then, these functions turn into zero for $\vec{\rho}^{\;2}\rightarrow\infty$
faster than $(\vec{\rho}^{\;2})^{-1}$ to provide the infrared safety. The
ultraviolet singularities of these functions at $\vec{\rho}=\vec{r}_{2}$\ and
$\vec{\rho}=\vec{r}_{1}$ cancel in the sum of the first three contributions in
the R.H.S. of (\ref{kernel through g}) on account of the \textquotedblleft
dipole\textquotedblright\ property of the \textquotedblleft
target\textquotedblright\ impact factors.

The last term in the R.H.S. of (\ref{kernel through g}) is the most complicated one:
\[
g(\vec{r}_{1},\vec{r}_{2};\vec{r}_{1}^{\;\prime},\vec{r}_{2}^{\;\prime
})=\left[  \frac{\left(  \vec{r}_{22^{\prime}}\,\,\vec{r}_{12}\right)  }
{\vec{r}_{11^{\prime}}^{\;2}\vec{r}_{22^{\prime}}^{\;2}\vec{r}_{1^{\prime
}2^{\prime}}^{\;2}}-\frac{2\left(  \vec{r}_{22^{\prime}}\,\,\vec
{r}_{11^{\prime}}\right)  }{\vec{r}_{11^{\prime}}^{\;2}\vec{r}_{22^{\prime}
}^{\;2}\vec{r}_{1^{\prime}2^{\prime}}^{\;2}}+\frac{2\left(  \vec
{r}_{22^{\prime}}\,\,\vec{r}_{12^{\prime}}\right)  \left(  \vec{r}
_{11^{\prime}}\,\,\vec{r}_{12^{\prime}}\right)  }{\vec{r}_{11^{\prime}}
^{\;2}\vec{r}_{22^{\prime}}^{\;2}\vec{r}_{1^{\prime}2^{\prime}}^{\;2}\vec
{r}_{12^{\prime}}^{\;2}}\right]  \ln\left(  \frac{\vec{r}_{12^{\prime}}^{\;2}
}{\vec{r}_{1^{\prime}2^{\prime}}^{\;2}}\right)
\]
\[
+\frac{1}{2\vec{r}_{1^{\prime}2^{\prime}}^{\;2}}\left[  \frac{(\vec
{r}_{11^{\prime}}\,\vec{r}_{22^{\prime}})}{\vec{r}_{11^{\prime}}^{\,\,2}
\,\vec{r}_{22^{\prime}}^{\,\,2}}+\frac{(\vec{r}_{21^{\prime}}\,\vec
{r}_{12^{\prime}})}{\vec{r}_{21^{\prime}}^{\,\,2}\,\vec{r}_{12^{\prime}
}^{\,\,2}}-\frac{2(\vec{r}_{22^{\prime}}\,\,\vec{r}_{21^{\prime}})}{\vec
{r}_{22^{\prime}}^{\;2}\vec{r}_{21^{\prime}}^{\;2}}\right]  \ln\left(
\frac{\vec{r}_{11^{\prime}}^{\,\,2}\,\vec{r}_{22^{\prime}}^{\,\,2}}{\vec
{r}_{1^{\prime}2^{\prime}}^{\,\,2}\vec{r}_{12}^{\,\,2}}\right)  +\frac
{(\vec{r}_{11^{\prime}}\,\vec{r}_{22^{\prime}})}{2\vec{r}_{11^{\prime}
}^{\,\,2}\,\vec{r}_{22^{\prime}}^{\,\,2}\vec{r}_{1^{\prime}2^{\prime}}^{\;2}
}\ln\left(  \frac{\vec{r}_{21^{\prime}}^{\,\,2}\,\vec{r}_{12^{\prime}}
^{\,\,2}}{\vec{r}_{11^{\prime}}^{\,\,2}\,\vec{r}_{22^{\prime}}^{\,\,2}
}\right)
\]
\[
+\frac{1}{d\,\vec{r}_{1^{\prime}2^{\prime}}^{\;2}}\left[  \frac{(\vec
{r}_{1^{\prime}2^{\prime}}\,\vec{r}_{12^{\prime}})\vec{r}_{12}^{\;2}}{\vec
{r}_{11^{\prime}}^{\;2}}+\frac{2(\vec{r}_{22^{\prime}}\,\vec{r}_{21^{\prime}
})(\vec{r}_{12}\,\vec{r}_{21^{\prime}})}{\vec{r}_{21^{\prime}}^{\;2}}
+\frac{(\vec{r}_{22^{\prime}}\,\vec{r}_{12^{\prime}})(\vec{r}_{11^{\prime}
}\,\vec{r}_{21^{\prime}})}{\vec{r}_{11^{\prime}}^{\;2}\,\vec{r}_{22^{\prime}
}^{\;2}}\vec{r}_{1^{\prime}2^{\prime}}^{\;2}-\vec{r}_{1^{\prime}2^{\prime}
}^{\;2}\right]  \ln\left(  \frac{\vec{r}_{12^{\prime}}^{\;2}\,\vec
{r}_{21^{\prime}}^{\;2}}{\vec{r}_{11^{\prime}}^{\;2}\,\vec{r}_{22^{\prime}
}^{\;2}}\right)
\]
\[
+\frac{1}{2\vec{r}_{1^{\prime}2^{\prime}}^{\,\,4}}\left(  \frac{\vec
{r}_{11^{\prime}}^{\;2}\,\vec{r}_{22^{\prime}}^{\;2}}{d}\ln\left(  \frac
{\vec{r}_{12^{\prime}}^{\;2}\,\vec{r}_{21^{\prime}}^{\;2}}{\vec{r}
_{11^{\prime}}^{\;2}\vec{r}_{22^{\prime}}^{\;2}}\right)  -1\right)  +\frac
{1}{\vec{r}_{11^{\prime}}^{\;2}\,}\left(  \frac{(\vec{r}_{12}\,\vec
{r}_{21^{\prime}})}{\vec{r}_{1^{\prime}2^{\prime}}^{\;2}\vec{r}_{21^{\prime}
}^{\;2}}-\frac{(\vec{r}_{11^{\prime}}\,\vec{r}_{12})}{\vec{r}_{1^{\prime
}2^{\prime}}^{\;2}\vec{r}_{22^{\prime}}^{\;2}}-\frac{(\vec{r}_{11^{\prime}
}\,\vec{r}_{21^{\prime}})}{\vec{r}_{22^{\prime}}^{\;2}\vec{r}_{21^{\prime}
}^{\;2}}\right)  \ln\left(  \frac{\vec{r}_{12^{\prime}}^{\;2}\,}{\vec
{r}_{11^{\prime}}^{\;2}}\right)
\]
\begin{equation}
-\frac{(\vec{r}_{12}\,\,\vec{r}_{22^{\prime}})}{\vec{r}_{1^{\prime}2^{\prime}
}^{\;2}\vec{r}_{22^{\prime}}^{\;2}\vec{r}_{12^{\prime}}^{\;2}}\ln\left(
\frac{\vec{r}_{11^{\prime}}^{\;2}}{\vec{r}_{1^{\prime}2^{\prime}}^{\;2}
}\right)  +(1\leftrightarrow2),
\end{equation}
\ where $d=\vec{r}_{12^{\prime}}^{\;2}\vec{r}_{21^{\prime}}^{\;2}-\vec
{r}_{11^{\prime}}^{\;2}\vec{r}_{22^{\prime}}^{\,\,2}.$

This term also vanishes at $\vec{r}_{1}=\vec{r}_{2}$, so that it possesses the
\textquotedblleft dipole\textquotedblright\ property. It has ultraviolet
singularity only at $\vec{r}_{1^{\prime}2^{\prime}}=0$ and tends to zero at
large $\vec{r}_{1}^{\;\prime\;2}$ and $\vec{r}_{2}^{\;\prime\;2}$ sufficiently
quickly in order to provide the infrared safety.

\section{Conclusion}

\label{sec:conclusion}

The coordinate representation of the BFKL kernel is extremely
interesting, because it gives the possibility to understand its
conformal properties and the relation between the BFKL and the
color dipole approaches. Generally speaking, the BFKL kernel is
not equivalent to the dipole one. Actually the first one is more
general than the second. This is clear, because the BFKL kernel
can be applied not only in the case of scattering of colourless
objects. However, when applied to the latter case, we can use the
\textquotedblleft dipole\textquotedblright\ and \textquotedblleft
gauge invariance\textquotedblright\ properties of targets and
projectiles and omit the terms in the kernel proportional to
$\delta(\vec{r}_{1^{\prime }2^{\prime}})$, as well as change the
terms independent either of $\vec{r} _{1}$ or of $\vec{r}_{2}$ in
such a way that the resulting kernel becomes conserving the
\textquotedblleft dipole\textquotedblright\ property, i.e. the
property which provides the vanishing of cross-sections for
scattering of zero-size dipoles.  The coordinate representation of
the kernel obtained in such a way is what we call the dipole form
of the BFKL kernel. We have found the dipole form of the gluon
contribution to the BFKL kernel in the NLO by the transfer of the
kernel from the momentum representation where it was calculated
before.  This paper completes the transformation of the NLO BFKL
kernel to the dipole form, started a few months ago with the quark
part of the kernel~\cite{Fadin:2006ha,Fadin:2007ee}.

The striking result of \cite{Fadin:2006ha, Fadin:2007ee} was the simplicity of
the dipole form of the quark contribution to the kernel. Moreover, it was
shown that the dipole form agrees with the quark contribution to the BK kernel
obtained in \cite{Balitsky:2006wa}.

As it can be seen from the results of this paper, the dipole form of
the gluon contribution to the kernel is also extremely simple in
the coordinate representation. This holds especially for the
\textquotedblleft symmetric\textquotedblright\ part of the kernel,
the momentum representation of which has a very complicated
form~\cite{FF05}.

We do not have the possibility to compare the results obtained for the NLO
gluon contribution with the BK kernel, since the latter has not been obtained yet. Instead,
our results can be used for finding the NLO BK kernel. In this respect, one has to bear
in mind the ambiguity of the NLO kernel related to the operator transformation
(\ref{transformation at NLO}) with an appropriate $\hat{O}$.

Note that in contrast to the LO, as well as to the quark contribution, the
functions $g^{0}(\vec{r}_{1},\vec{r}_{2};\vec{\rho})$ and $g(\vec{r}_{1}
,\vec{r}_{2};\vec{\rho})$ in the dipole form (\ref{kernel through g}) turned
out to be unequal. Although the function $g^{0}(\vec{r}_{1},\vec{r}_{2}
;\vec{\rho})$ can be changed by adding any function with zero
integral over $\vec{\rho}$, the inequality cannot be removed. On
the other hand, according to~\cite{Balitsky:2006wa} in the colour
dipole approach these functions should be equal.

We have to say that our consideration was not completely rigorous.
In particular, we did not regularize the ultraviolet singularities
arising as a result of separation of the ultraviolet non-singular
sum $V(\vec{k})+V(\vec{k} ,\vec{k}-\vec{q})$ in (\ref{planar part
for integration}) into two pieces. Instead of the regularization
we used the trick  of representing the coefficient of
$\delta(\vec{r}_{11^{\prime}})\delta(\vec{r}_{22^{\prime}})$ in
(\ref{kernel through g}) in integral form.  Note, however,
that the trick is the same which was used in
Ref.~\cite{Fadin:2006ha}, where its validity was checked by
completely rigorous calculations. This permits to rely on the
results obtained here.

\vspace{0.5cm} {\textbf{\Large Acknowledgments}} \vspace{0.5cm}

We would like to thank R.E. Gerasimov for calculating the integral~(\ref{int f4}).

\section{Appendix A}

\label{sec:appendix A} Here we present a list of the integrals necessary to
perform the Fourier transform of the planar part of the kernel. Most of them are
calculated straightforwardly via the exponential representation:
\begin{equation}
a^{-j}=\frac{1}{\Gamma(j)}\int_{0}^{\infty}d\alpha\;\alpha^{j-1}e^{-\alpha a}.
\end{equation}
We have
\begin{equation}
\int\frac{d\vec{k}}{2\pi}e^{i\vec{k}\,\vec{r}}\frac{\vec{k}}{\vec{k}^{\,2}
}=\frac{i\vec{r}}{\vec{r}^{\,2}},\qquad\int\frac{d\vec{k}}{2\pi}e^{i\vec
{k}\,\vec{r}}\ln\vec{k}^{2}=-\frac{2}{\vec{r}^{\,2}}, \label{int k/k^2}
\end{equation}
\begin{equation}
\int\frac{d\vec{k}}{2\pi}e^{i\vec{k}\,\vec{r}}\frac{\vec{k}}{\vec{k}^{\,2}}
\ln\frac{\vec{k}^{\,2}}{\mu^{2}}=\frac{i\vec{r}}{\vec{r}^{\,2}}\left(
2\psi(1)-\ln\left(  \frac{\vec{r}^{\,2}\mu^{2}}{4}\right)  \right)  ,
\label{int k/k^2 ln k/k^2}
\end{equation}
\begin{equation}
\int\frac{d\vec{q}}{2\pi}\int\frac{d\vec{k}}{2\pi}e^{i[\vec{q}\,\vec{r}
+\vec{k}\,\vec{\rho}]}\frac{(\vec{q}\,\vec{k})}{\vec{q}^{\,2}\vec{k}^{\,2}}
\ln\frac{(\vec{k}+\vec{q})^{2}}{\vec{q}^{\,2}}=-\frac{\left(  \vec{r}
\,\vec{\rho}\right)  }{\vec{r}^{\,2}\,\vec{\rho}^{\,2}}\ln\left(  \frac
{(\vec{\rho}-\vec{r})^{2}}{\vec{\rho}^{\,2}}\right)  ,
\label{int (qk)/qk ln(k+q)/q}
\end{equation}
\begin{equation}
\int\frac{d\vec{q}}{2\pi}\int\frac{d\vec{k}}{2\pi}e^{i[\vec{q}\,\vec{r}
+\vec{k}\,\vec{\rho}]}\frac{(\vec{q}\,\vec{k})}{\vec{q}^{\,2}\vec{k}^{\,2}}
\ln^{2}\frac{(\vec{k}+\vec{q})^{2}}{\vec{q}^{\,2}}=-\frac{\left(  \vec
{r}\,\vec{\rho}\right)  }{\vec{r}^{\,2}\vec{\rho}^{\,2}}\ln^{2}\left(
\frac{(\vec{\rho}-\vec{r})^{2}}{\vec{\rho}^{\,2}}\right)  ,
\end{equation}
\begin{equation}
\int\frac{d\vec{q}}{2\pi}\int\frac{d\vec{k}}{2\pi}e^{i[\vec{q}\,\vec{r}
+\vec{k}\,\vec{\rho}]}\frac{(\vec{q}\,\vec{k})}{\vec{q}^{\,2}\vec{k}^{\,2}}
\ln^{2}\frac{\vec{k}^{\,2}}{\vec{q}^{\,2}}=-\frac{\left(  \vec{r}\,\vec{\rho
}\right)  }{\vec{r}^{\,2}\vec{\rho}^{\,2}}\ln^{2}\left(  \frac{\vec{\rho
}^{\,2}}{\vec{r}^{\,2}}\right)  ,
\end{equation}
\begin{align}
&  \int\frac{d\vec{q}}{2\pi}\int\frac{d\vec{k}}{2\pi}e^{i[\vec{q}\,\vec
{r}+\vec{k}\,\vec{\rho}]}\frac{1}{\vec{q}^{\,2}}\ln^{2}\frac{(\vec{q}-\vec
{k})^{2}}{\vec{k}^{\,2}}=\frac{1}{\vec{\rho}^{\,2}}\ln^{2}\left(
\frac{\left(  \vec{r}+\vec{\rho}\right)  ^{2}}{\vec{r}^{\,2}}\right)  ,\\
&  \int\frac{d\vec{q}}{2\pi}\int\frac{d\vec{k}}{2\pi}e^{i[\vec{q}\,\vec
{r}+\vec{k}\,\vec{\rho}]}\frac{1}{\vec{q}^{\,2}}\ln\frac{(\vec{q}-\vec{k}
)^{2}}{\vec{k}^{\,2}}=\frac{1}{\vec{\rho}^{\,2}}\ln\left(  \frac{\left(
\vec{r}+\vec{\rho}\right)  ^{2}}{\vec{r}^{\,2}}\right)  ,
\label{int 1/q ln(k+q)/k}
\end{align}
\begin{equation}
\int\frac{d\vec{q}}{2\pi}\int\frac{d\vec{k}}{2\pi}e^{i[\vec{q}\,\vec{r}
+\vec{k}\,\vec{\rho}]}\frac{1}{\vec{q}^{\,2}}\ln\frac{(\vec{q}-\vec{k})^{2}
}{\vec{k}^{\,2}}\ln\frac{(\vec{q}-\vec{k})^{2}}{\vec{q}^{\,2}}=\frac{1}
{\vec{\rho}^{\,2}}\ln\left(  \frac{\left(  \vec{r}+\vec{\rho}\right)  ^{2}
}{\vec{r}^{\,2}}\right)  \ln\left(  \frac{\left(  \vec{r}+\vec{\rho}\right)
^{2}}{\vec{\rho}^{\,2}}\right)  . \label{int 1/q ln(q+k)/q ln(q+k)/k}
\end{equation}
The integral (\ref{int 1/q ln(q+k)/q ln(q+k)/k}) is calculated most easily after
the decomposition $\ln a\ln b=\frac{1}{2}(\ln^{2}a+\ln^{2}b-\ln^{2}\frac{a}
{b}).$
\begin{equation}
\int\frac{d\vec{q}_{1}}{2\pi}\int\frac{d\vec{q}_{2}}{2\pi}\int\frac{d\vec{k}
}{2\pi}e^{i[\vec{q}_{1}\,\vec{r}_{1}+\vec{q}_{2}\,\vec{r}_{2}+\vec{k}
\,\vec{\rho}]}\frac{(\vec{q}_{1}\,\vec{q}_{2})}{\vec{q}_{1}^{\;2}\vec{q}
_{2}^{\;2}}\ln^{2}\frac{(\vec{q}_{1}+\vec{q}_{2})^{2}}{\vec{k}^{\,2}}
=\frac{4(\vec{r}_{1}\,\vec{r}_{2})}{\vec{r}_{1}^{\,2}\vec{r}_{2}^{\,2}
\vec{\rho}^{\,2}}\ln\left(  \frac{\vec{r}_{1}^{\,2}\vec{r}_{2}^{\,2}}{\left(
\vec{r}_{1}-\vec{r}_{2}\right)  ^{2}\vec{\rho}^{\,2}}\right)  ,
\label{int (q1q2)/q1q2 ln(q1+q2)/k}
\end{equation}
\begin{equation}
\int\frac{d\vec{q}_{1}}{2\pi}\int\frac{d\vec{q}_{2}}{2\pi}\int\frac{d\vec{k}
}{2\pi}e^{i[\vec{q}_{1}\,\vec{r}_{1}+\vec{q}_{2}\,\vec{r}_{2}+\vec{k}
\,\vec{\rho}]}\frac{(\vec{q}_{1}\,\vec{q}_{2})}{\vec{q}_{1}^{\;2}\vec{q}
_{2}^{\;2}}\ln^{2}\frac{\vec{q}_{1}^{\;2}}{\vec{k}^{\,2}}=\frac{4(\vec{r}
_{1}\,\vec{r}_{2})}{\vec{r}_{1}^{\,2}\vec{r}_{2}^{\,2}\vec{\rho}^{\,2}}
\ln\left(  \frac{\vec{r}_{1}^{\,2}}{\vec{\rho}^{\,2}}\right)  .
\label{int (q1q2)/q1q2 ln q1/k}
\end{equation}
The following integrals contain $I$ defined in (\ref{integral I}):
\begin{align}
\int\frac{d\vec{q}}{2\pi}\int\frac{d\vec{k}}{2\pi}e^{i[\vec{q}\,\vec{r}
+\vec{k}\,\vec{\rho}]}I(\vec{q}^{\;2},(\vec{q}-\vec{k})^{2},\vec{k}^{\;2})  &
=I(\vec{\rho}^{\;2},(\vec{\rho}+\vec{r})^{2},\vec{r}^{\;2}
),\label{int I(k,q,k+q)}\\
\int\frac{d\vec{q}}{2\pi}\int\frac{d\vec{k}}{2\pi}e^{i[\vec{q}\,\vec{r}
+\vec{k}\,\vec{\rho}]}\frac{(\vec{q}\,\vec{k})}{\vec{k}^{\,\,2}}I(\vec
{q}^{\;2},(\vec{q}-\vec{k})^{2},\vec{k}^{\;2})  &  =-\frac{(\vec{r}\,\vec
{\rho})}{\vec{r}^{\;2}}I(\vec{\rho}^{\;2},(\vec{\rho}+\vec{r})^{2},\vec
{r}^{\;2}), \label{int (qk)/k I(q,k,q+k)}
\end{align}
\begin{equation}
\int\frac{d\vec{q}}{2\pi}\int\frac{d\vec{k}}{2\pi}e^{i[\vec{q}\,\vec{r}
+\vec{k}\,\vec{\rho}]}\frac{(\vec{q}\,\,\vec{k})^{2}}{\vec{q}^{\,2}\vec
{k}^{\;2}}I(\vec{q}^{\;2},(\vec{q}-\vec{k})^{2},\vec{k}^{\;2})=\frac{(\vec
{r}\,\vec{\rho})^{2}}{\vec{r}^{\;2}\vec{\rho}^{\;2}}I(\vec{\rho}^{\;2}
,(\vec{\rho}+\vec{r})^{2},\vec{r}^{\;2}). \label{int (kq)/kq I(q,k,k+q)}
\end{equation}
This integral can be expressed through (\ref{int (qk)/k I(q,k,q+k)}) and
simpler integrals with the help of the identity
\begin{equation}
(\vec{k}\,\vec{q})=\frac{(1-x(1-z))\vec{k}^{\;2}}{2(1-x)(1-z)}+\frac
{(z+x(1-z))\vec{q}^{\;2}}{2x(1-z)}-\frac{(\vec{q}-\vec{k})^{2}
x(1-x)(1-z)+(\vec{q}^{\;2}(1-x)+\vec{k}^{\;2}x)z}{2(1-x)x(1-z)}.
\end{equation}

\section{Appendix B}

\label{sec:appendix B}

Here we present a list of the integrals necessary to perform the Fourier transform of
the symmetric part of the kernel:
\[
\int\frac{d\vec{q}}{2\pi}\int\frac{d\vec{p}}{2\pi}e^{i[\vec{p}\,\,\vec{r}
+\vec{q}\,\vec{\rho}]}\frac{\vec{p}}{\vec{q}^{\,2}\vec{p}^{\;2}}\ln\frac
{\vec{p}^{\;2}}{(\vec{q}+\vec{p})^{\,2}}=-\frac{i}{2}\left[  \vec{\rho
}\,I(\vec{\rho}^{\,2},\vec{r}^{\,2},(\vec{r}-\vec{\rho})^{2})\right.
\]
\begin{equation}
+\left.  \frac{\vec{r}}{\vec{r}^{\,2}}\left(  -(\vec{r}\,\vec{\rho}
)I(\vec{\rho}^{\,2},\vec{r}^{\,2},(\vec{r}-\vec{\rho})^{2})+\frac{1}{2}
\ln\left(  \frac{\vec{\rho}^{\,2}}{(\vec{\rho}-\vec{r})^{2}}\right)
\ln\left(  \frac{\vec{\rho}^{\,2}}{\vec{r}^{\,2}}\right)  \right)  \right]  .
\label{int p/qp ln p/(q+p)}
\end{equation}
\begin{equation}
\int\frac{d\vec{q}}{2\pi}\int\frac{d\vec{k}}{2\pi}e^{i\vec{q}\,\vec{\rho}
}\left(  e^{i\vec{k}\,\vec{r}_{1}}-e^{i\vec{k}\,\vec{r}_{2}}\right)
\frac{\vec{q}}{\vec{q}^{\;2}\vec{k}^{\;2}}\ln\frac{\vec{q}^{\;2}}{\vec{k}^{\;2}
}=\frac{i\vec{\rho}}{4\vec{\rho}^{\;2}}\ln\left(  \frac{\vec{r}_{1}^{\;2}
}{\vec{r}_{2}^{\;2}}\right)  \ln\left(  \frac{\vec{\rho}^{\;4}}{\vec{r}
_{1}^{\;2}\vec{r}_{2}^{\;2}}\right)  , \label{int (exp-exp)q/qk ln q/k}
\end{equation}
\begin{equation}
\int\frac{d\vec{q}}{2\pi}\frac{d\,\vec{l}}{2\pi}\frac{e^{i[\vec{q}\,\vec
{r}+\vec{l}\,\vec{\rho}]}}{\vec{l}^{\;2}+x(1-x)\vec{q}^{\;2}}=\frac{1}{\vec
{r}^{\;2}+x(1-x)\vec{\rho}^{\;2}}\;, \label{int 1/sigma}
\end{equation}
\begin{equation}
\int\frac{d\vec{q}}{2\pi}\frac{d\,\vec{l}}{2\pi}\frac{q_{i}l_{j}}{\vec
{q}^{\;2}}\frac{e^{i[\vec{q}\,\vec{r}+\vec{l}\,\vec{\rho}]}}{\vec{l}
^{\;2}+x(1-x)\vec{q}^{\;2}}=\frac{-r_{i}\rho_{j}}{\vec{\rho}^{\;2}(\vec
{r}^{\;2}+x(1-x)\vec{\rho}^{\;2})}\;, \label{int (ql)/(q sigma)}
\end{equation}
\begin{equation}
\int\frac{d\vec{q}}{2\pi}\frac{d\vec{k}}{2\pi}\frac{q_{i}k_{j}}{\vec{k}^{\;2}
}\frac{e^{i[\vec{q}\vec{r}+\vec{k}\vec{\rho}]}}{(1-x)\vec{k}^{\;2}+x\vec
{q}^{\;2}}=\frac{-r_{i}\rho_{j}}{\vec{r}^{\;2}((1-x)\vec{r}^{\;2}+x\vec{\rho
}^{\;2})}, \label{int (qk)/(k sigma)}
\end{equation}
\begin{equation}
\int\frac{d\vec{k}}{2\pi}\left(  e^{i\vec{k}\,\vec{r}_{1}}-e^{i\vec{k}
\,\vec{r}_{2}}\right)  \frac{1}{\vec{k}^{\;2}}=\frac{1}{2}\ln\left(
\frac{\vec{r}_{2}^{\;2}}{\vec{r}_{1}^{\;2}}\right)  . \label{int (exp-exp)1/k}
\end{equation}
\begin{equation}
\int\frac{d\vec{q}}{2\pi}\int\frac{d\vec{k}}{2\pi}e^{i[\vec{q}\vec{r}+\vec
{k}\vec{\rho}]}\frac{(\vec{q}\,\,\vec{k})}{\vec{q}^{\;2}\vec{k}^{\;2}}
\frac{(\vec{q}+\vec{k})}{(\vec{q}+\vec{k})^{2}}=-\frac{i}{4}\left(  \frac
{\vec{r}}{\vec{r}^{\;2}}\ln\frac{\vec{\rho}^{\;2}}{(\vec{r}-\vec{\rho})^{2}
}+\frac{\vec{\rho}}{\vec{\rho}^{\;2}}\ln\frac{\vec{r}^{\;2}}{(\vec{r}-\vec
{\rho})^{2}}\right)  \;, \label{int f0}
\end{equation}
For calculating the following integrals we use the identities
\begin{equation}
\frac{\partial I(a,b,c)}{\partial a}=\frac{1}{(c-a-b)^{2}-4ab}\left(
(c+b-a)I(a,b,c)+2\ln\frac{a}{c}+\frac{\left(  c-a-b\right)  }{a}\ln\frac{b}
{c}\right)  , \label{identity_with_derivative}
\end{equation}
\begin{align}
&  \int_{0}^{1}\frac{2\,c\,x\,dx}{ax+b(1-x)-cx(1-x)}\ln\left(  \frac
{ax+b(1-x)}{cx(1-x)}\right) \nonumber\\
&  =I(a,b,c)(c+b-a)+\mathrm{Li}_{2}\left(  1-\frac{b}{a}\right)
-\mathrm{Li}_{2}\left(  1-\frac{a}{b}\right)  +\frac{1}{2}\ln\frac{a}{b}
\ln\frac{ab}{c^{2}},
\end{align}
where
\begin{equation}
\mathrm{Li}_{2}\left(  x\right)  =-\int_{0}^{1}\frac{dt\,\ln(1-xt)}{t}.
\end{equation}
\[
\int\frac{d\vec{q}}{2\pi}\int\frac{d\vec{k}}{2\pi}e^{i[\vec{q}\vec{r}+\vec
{k}\vec{\rho}]}\frac{(\vec{q}\,\,\vec{k})}{\vec{q}^{\;2}\vec{k}^{\;2}}
\frac{(\vec{q}+\vec{k})}{(\vec{q}+\vec{k})^{2}}\ln\vec{k}^{\;2}=\frac{i}
{4}\left\{  \frac{\vec{r}}{\vec{r}^{\;2}}\left(  \frac{1}{2}\ln^{2}\left(
\frac{\vec{\rho}^{\;2}}{(\vec{r}-\vec{\rho})^{2}}\right)  +\vec{r}^{\;2}
I(\vec{r}^{\;2},\vec{\rho}^{\;2},(\vec{r}-\vec{\rho})^{2})\right)  \right.
\]
\[
+\left(  \frac{\vec{\rho}}{\vec{\rho}^{\;2}}\ln\left(  \frac{\vec{r}^{\;2}
}{(\vec{r}-\vec{\rho})^{2}}\right)  +\frac{\vec{r}}{\vec{r}^{\;2}}\ln\left(
\frac{\vec{\rho}^{\;2}}{(\vec{r}-\vec{\rho})^{2}}\right)  \right)  \left(
-2\psi(1)+\ln\frac{(\vec{r}-\vec{\rho})^{2}}{4}\right)
\]

\begin{equation}
+\left.  \frac{\vec{\rho}}{\vec{\rho}^{\;2}}\left(  \frac{1}{2}\ln\left(
\frac{\vec{\rho}^{\;2}}{(\vec{r}-\vec{\rho})^{2}}\right)  \ln\left(
\frac{\vec{r}^{\;2}}{(\vec{r}-\vec{\rho})^{2}}\right)  -(\vec{\rho}\,\,\vec
{r})I(\vec{r}^{\;2},\vec{\rho}^{\;2},(\vec{r}-\vec{\rho})^{2})\right)
\right\}  ,\label{int f1}
\end{equation}
\begin{align}
&  \int\frac{d^{2}q}{2\pi}\int\frac{d^{2}k}{2\pi}e^{i[\vec{q}(\vec{r}
-\vec{\rho})-\vec{k}\vec{\rho}]}\frac{q^{i}}{\vec{q}^{\;2}}\frac{\left(
q+k\right)  ^{j}}{(\vec{q}+\vec{k})^{2}}\ln\vec{k}^{\;2}\nonumber\\
&  =\frac{1}{4\vec{r}^{\,\,2}\vec{\rho}^{\,\,2}}\left(  \delta^{ij}(\vec
{r}\,\,\vec{\rho})+\rho^{j}r^{i}-\rho^{i}r^{j}\right)  \ln\left(  \frac
{(\vec{r}-\vec{\rho})^{4}}{\vec{r}^{\;2}\vec{\rho}^{\;2}}\right)  +\frac
{\rho^{j}r^{i}}{2\vec{r}^{\;2}\vec{\rho}^{\;2}}\left(  4\psi(1)-\ln\left(
\frac{\vec{\rho}^{\;2}\vec{r}^{\;2}}{16}\right)  \right)  \nonumber\\
&  +\frac{1}{2(\vec{r}-\vec{\rho})^{2}}\left(  \frac{\rho^{i}\rho^{j}}
{\vec{\rho}^{\;2}}-\frac{r^{i}r^{j}}{\vec{r}^{\;2}}+\frac{\left(  \vec{r}
^{\;2}-\vec{\rho}^{\;2}\right)  }{2\vec{r}^{\;2}\vec{\rho}^{\;2}}\left(
\delta^{ij}(\vec{r}\,\,\vec{\rho})-\rho^{j}r^{i}-\rho^{i}r^{j}\right)
\right)  \ln\left(  \frac{\vec{\rho}^{\;2}}{\vec{r}^{\;2}}\right)
.\label{int f2}
\end{align}
We simplified the tensor structure of this integral via the identity
\begin{equation}
\delta^{ij}=\frac{\rho^{i}\rho^{j}\vec{r}^{\;2}+r^{i}r^{j}\vec{\rho}
^{\;2}-(\vec{r}\,\,\vec{\rho})\left(  \rho^{j}r^{i}+\rho^{i}r^{j}\right)
}{\vec{r}^{\,\,2}\vec{\rho}^{\,\,2}-(\vec{r}\,\,\vec{\rho})^{2}}.
\end{equation}
The integral
\begin{equation}
J=\int\frac{d\vec{q}_{1}}{2\pi}
\int\frac{d\vec{q}_{2}}{2\pi}\int\frac{d\vec{k}}{2\pi}e^{i[\vec{q}_{1}
\,\vec{r}_{11^{\prime}}+\vec{q}_{2}\,\vec{r}_{22^{\prime}}+\vec{k}\,\vec
{r}_{1^{\prime}2^{\prime}}]}\frac{(\vec{q}_{1}\,\vec{q}_{2})}{\vec{q}
_{1}^{\;2}\vec{q}_{2}^{\;2}}\ln\frac{(\vec{q}_{1}-\vec{k})^{2}}{(\vec{q}
_{1}+\vec{q}_{2})^{2}}\ln\frac{(\vec{q}_{2}+\vec{k})^{2}}{(\vec{q}_{1}+\vec
{q}_{2})^{2}} \label{integral J}
\end{equation}
appearing in the  \textquotedblleft planar\textquotedblright part
can be written as
\begin{equation}
J=\frac{8}{r_{1^{\prime}2^{\prime}}^{2}}\int\frac{d\vec{\rho}}{2\pi}\frac{\left(  \vec{r}_{1\rho}\,\,\vec{r}
_{2\rho}\right)
}{\vec{r}_{1\rho}^{\;2}\vec{r}_{2\rho}^{\;2}}\frac{(\vec
{r}_{1^{\prime}\rho}\,\,\vec{r}_{2^{\prime}\rho})}{\vec{r}_{1^{\prime}\rho
}^{\;2}\vec{r}_{2^{\prime}\rho}^{\;2}}\label{int f4 before integration}
\end{equation}
putting
\begin{equation}
\int\frac{d\vec{\rho}}{\left(  2\pi\right)  ^{2}}d\vec{q}_{1}^{\;\prime}
d\vec{q}_{2}^{\;\prime}\delta\left(  \vec{q}_{1}-\vec{k}-\vec{q}_{1}
^{\;\prime}\right)  e^{i\,\vec{r}_{2^{\prime}\rho}[\vec{q}_{1}+\vec{q}
_{2}-\vec{q}_{1}^{\;\prime}-\vec{q}_{2}^{\;\prime}]}=1~
\end{equation}
into the integrand in (\ref{integral J}) and then integrating over
momenta with the help of (\ref{int k/k^2}). Using the same trick
we obtain
\begin{equation}
\int\frac{d\vec{q}_{1}}{2\pi}\int\frac{d\vec{q}_{2}}{2\pi}\int\frac
{d\vec{k}}{2\pi}e^{i[\vec{q}_{1}\,\vec{r}_{11^{\prime}}+\vec{q}_{2}\,\vec
{r}_{22^{\prime}}+\vec{k}\,\vec{r}_{1^{\prime}2^{\prime}}]}\frac{\left(
\vec{q}_{1}\,\,\vec{q}_{2}\right)  }{\vec{q}_{1}^{\;2}\vec{q}_{2}^{\;2}}
\frac{(\vec{q}_{1}-\vec{k})}{(\vec{q}_{1}-\vec{k})^{2}}\frac{(\vec{q}_{2}
+\vec{k})}{(\vec{q}_{2}+\vec{k})^{2}}=\frac{r_{1^{\prime}2^{\prime}}^{2}}{8}J
\end{equation}
for the integral appearing in the  \textquotedblleft
symmetric\textquotedblright\ part.  One can calculate $J$\ by
means of complex variables
\[
a_{+}=a_{x}+i\,a_{y},\quad a_{-}=a_{x}-i\,a_{y},\quad\left(  \vec{a}
\,\,\vec{c}\right)  =\frac{a_{+}c_{-}+a_{-}c_{+}}{2},\quad\vec{a}
^{\,\,\,2}=a_{+}a_{-}.
\]
Shifting $\vec{\rho}\rightarrow\vec{r}_{1}-\vec{\rho},$\ introducing
$z=e^{i\phi}$, using
\[
\rho_{+}=\rho z,\quad\rho_{-}=\frac{\rho}{z},\quad d\phi=\frac{d\,z}{i\,z}
\]
and integrating over $z$\ via residues, one gets only trivial
integrals over $\rho$\ to perform. They yield
\begin{equation}
J=\frac{2}{\vec r_{1^{\prime}2^{\prime}}^{\,\,2}}\left(\frac{(\vec{r}_{11^{\prime}}\,
\vec{r}_{22^{\prime}})}{\vec{r}_{11^{\prime
}}^{\,\,2}\,\vec{r}_{22^{\prime}}^{\,\,2}}\ln\left(  \frac{\vec{r}
_{21^{\prime}}^{\,\,2}\,\vec{r}_{12^{\prime}}^{\,\,2}}{\vec{r}_{1^{\prime
}2^{\prime}}^{\,\,2}\vec{r}_{12}^{\,\,2}}\right)
+\frac{(\vec{r}_{21^{\prime
}}\,\vec{r}_{12^{\prime}})}{\vec{r}_{21^{\prime}}^{\,\,2}\,\vec
{r}_{12^{\prime}}^{\,\,2}}\ln\left(  \frac{\vec{r}_{11^{\prime}}^{\,\,2}
\,\vec{r}_{22^{\prime}}^{\,\,2}}{\vec{r}_{1^{\prime}2^{\prime}}^{\,\,2}\vec
{r}_{12}^{\,\,2}}\right)\right)  .\label{int f4}
\end{equation}
\

\end{document}